\documentclass[twocolumn]{aastex631}
\usepackage{CJK}
\usepackage{hyperref}

\newcommand{\oiiil}{[O\,{\footnotesize III}] $\lambda$5007}
\newcommand{\ocl}{[O\,{\footnotesize III}] $\lambda$4959}
\newcommand{\oiii}{[O\,{\footnotesize III}]}

\newcommand{\hb}{\rm H\ensuremath{\beta}}

\def\lephare{\texttt{LePhare}}
\newcommand{\ppxf}{\textsc{ppxf}}

\begin{document}
\begin{CJK*}{UTF8}{gbsn}

%\title{Outflows in IR selected AGNs\footnote{Released on March, 1st, 2021}}
\title{Host galaxy and nuclear properties of IR-selected AGNs with and without outflow signatures \footnote{Released on December, 20st, 2023}}

\correspondingauthor{Gabriel A. Oio}
\email{gabrieloio@unc.edu.ar}
\author[0000-0002-7938-6107]{Gabriel A. Oio}
\affiliation{Chinese Academy of Sciences South America Center for Astronomy (CASSACA), \\
National Astronomical Observatories of China (NAOC),\\
CAS, 20A Datun Road, Beijing 100012, China}

\correspondingauthor{Y. Sophia Dai (戴昱)}
\email{ydai@nao.cas.cn}
\author[0000-0002-7928-416X]{Y. Sophia Dai (戴昱)}
\affiliation{Chinese Academy of Sciences South America Center for Astronomy (CASSACA), \\
National Astronomical Observatories of China (NAOC),\\
CAS, 20A Datun Road, Beijing 100012, China}

\author[0000-0001-6800-3329]{C. G. Bornancini}
\affiliation{Instituto de Astronom\'ia Te\'orica y Experimental, (IATE, CONICET-UNC), C\'ordoba, Argentina}
\affiliation{Observatorio Astron\'omico de C\'ordoba, Universidad Nacional de C\'ordoba, Laprida 854, X5000BGR, C\'ordoba, Argentina}

\author[0000-0001-7634-1547]{Zi-Jian Li}
\affiliation{Chinese Academy of Sciences South America Center for Astronomy (CASSACA), \\
National Astronomical Observatories of China (NAOC),\\
CAS, 20A Datun Road, Beijing 100012, China}
\affiliation{National Astronomical Observatories, Chinese Academy of Sciences, 20A Datun Road, 100012 Beijing, China}
\affiliation{School of Astronomy and Space Sciences, University of Chinese Academy of Sciences, Beijing 100049, China}


\begin{abstract}
Active galactic nucleus (AGN) driven outflows can have a significant impact on the evolution of the host galaxy. In this work, we compare the properties of galaxies that hosts AGNs with and without outflows. Our sample consists of 103 AGNs identified by mid-IR color-color selection, and confirmed with optical spectroscopy at a redshift range of 0.3 $\lesssim$ z $\lesssim$ 0.9. We fit the \oiiil~ line using spectra from the zCOSMOS survey to identify and to study the occurrence of outflows. 
We find that ionized outflows are present in $\sim$25\% of our sample, with the largest incidence at the highest \oiii~ and X-ray luminosity bins. The fastest outflows are found in the more extended and massive galaxies. We do not observe a difference in the star formation rate of AGNs with outflows compared to AGNs without outflows.  From visual inspection and non-parametric morphological studies, we obtain that outflows are preferentially observed in galaxies with elliptical morphologies.

\end{abstract}



\section{Introduction} \label{sec:intro}
%Introduction to AGN feedback and outflows.\\
%introduction to AGN selection , esp. IR selection. \\
%introduction to outflow selection methods, includ. [OIII] selections.\\
%Physical mechanisms and previous works on the subject.\\
Active galactic nuclei (AGN) are intrinsically linked with the origin and evolution of galaxies. The relationships between black holes and various parameters related to their host galaxies are well known, but differ by orders of magnitude in their physical scale. Among them, we can name the relationship between black hole mass and bulge mass \citep{magorrian98,wandel,mclure,haring04,graham15,ding}, between black hole mass and velocity dispersion \citep{ferrarese00,Gebhardt00,merrit01,beifiori}, and even with the total mass of the host galaxy \citep{bandara09}. The co-evolution of these different components is believed to be driven by feedback \citep{silkrees98,dimatteo2005,Fabian2012,Harrison2018,Veilleux2020}.
This feedback regulates the accumulation of mass in galaxies by heating and/or removing the gas that would otherwise be used to form stars, and by ejecting matter as mass is added to the central black hole. %There are two main modes of AGN feedback: quasar or radiative mode and radio or kinetic mode \citep[see e.g.,][]{Fabian2012}. Quasar mode feedback occurs when the AGN is accreting matter at a high rate. This produces a powerful jet of plasma that can heat up and/or remove the gas in the host galaxy. Radio mode feedback occurs when the AGN is accreting matter at a lower rate. This produces a slower, wider-angle wind that can also heat up and/or remove gas from the host galaxy.
This process, where warm ionized and cold molecular outflows are being driven by an AGN accreting near the Eddington limit, is usually referred to as ``quasar mode" \citep[see e.g.,][]{Fabian2012,Harrison2018}.

There are numerous approaches in the literature to define outflows from the emission lines.
One way that has been used to find outflowing gas emission in AGNs is by considering the line-of-sight velocity offset ($v_{off}$) of narrow emission lines with respect to the systemic velocity of their host galaxies \citep[e.g.,][]{boroson2005,barbosa2009,Muller2011}. This approach is based on the idea that there is a stratified narrow line region (NLR) where, in its innermost regions, high ionization lines such as \oiii~are blueshifted and more turbulent (broader) than the low-ionization lines produced in the external emitting regions and also with respect to the stellar absorption lines that are indicative of the systemic velocity of their host galaxies. \citep[e.g.,][]{crenshaw2000,ruiz2005,komossa2008,bae2014,schmidt2018,trindade2021}.
Given that the forbidden \oiiil~emission lines cannot be produced in the high-density, sub-parsec scales broad line regions (BLR) of AGNs, they are usually considered to be a good tracer of the extended ionized gas. Particularly in AGNs, the \oiiil~line profile can show a broad, blue-wing asymmetry, which is usually attributed to an outflowing gas. Therefore, some works focus on the multi-component fitting of \oiiil~line and characterize the outflow by its blue-wing properties \citep[e.g.,][]{komossa2007,holt2008,mullaney2013,zakamska2014,carniani2015,Harrison2016,schmidt2018,Oio2019,guolo2021}. In contrast, some authors have chosen to characterize the outflow from its full line profile, using so-called non-parametric definitions \citep[e.g.,][]{Liu2013b,Harrison2014,zakamska2014,Lei2017, robleto2021}.
\cite{Liu2013b} measured the \oiii~line width containing the central 80 per cent of flux (W$_{80}$), in a sample of nearby, luminous, obscured, radio-quiet quasars. In their work, they used this line width as an estimate of the outflow velocity, and found that most objects showed a blue excess in their line profiles, with a median outflow velocity of 760 km s$^{-1}$. 
In a similar manner, \cite{Harrison2014} employed the same approach to measure the line width of the emission lines \ocl~and \oiiil. Their objective was to demonstrate the prevalence of outflows in a sample of 16 luminous type II AGNs. Here, they found extended \oiii~emission, with W$_{80}$ ranging from 600 to 1500 km s$^{-1}$.

Some results obtained in numerical simulations suggest that there is a stage in the evolution of galaxies where violent interactions and mergers are more frequent. In these evolutionary scenarios the first instances of mergers would show galaxies with an obscured nuclei by large amounts of gas and dust. Then a phase would follow where the AGN drives outflows that expel the surrounding material and reveal an unobscured AGN \citep{dimatteo2005,springel2005,hopkins08a, hopkins08b}.
The relationship between gas outflows or inflows in AGNs and the obscuration measured by optical spectral line widths or hard/soft X-ray emission is still not well understood.
We might expect that obscured AGNs selected in the X-rays, according to high neutral hydrogen column densities or high hardness ratio values, calculated as HR$=$H$-$S/H$+$S, where $H$ and $S$ are the count rates in the hard and soft bands, to also have outflow signatures \citep{Harrison2016}.
%However, the results obtained in the literature differ. 
\citet{Ricci2017} using the column density ($N_{\rm H}$) estimated from X-ray spectral analysis showed opposite results.
In  the same way \citet{Harrison2016} found no evidence that the highest ionized gas velocities are preferentially associated with X-ray obscured AGN.
\citet{Rojas2020} in a study of X-ray selected (z$\sim$0.05) AGN taken from the BAT AGN Spectroscopic Survey (BASS) found that the occurrence of outflow detections in unobscured type 1 and type 1.9 (with narrow lines except for a broad component seen in H$\alpha$) AGNs is twice that of obscured (type 2) AGNs.

Mergers and interactions have long been considered a triggering mechanism for AGN \citep{CanalizoStockton2001,Hopkins2008,RamosAlmeida2011,Bessiere2012,Satyapal2014,Goulding2018,Pierce2023}.
Because of the increase of nuclear star formation and nuclear activity in mergers \citep[e.g.,][]{Satyapal2014}, it is reasonable to consider that the incidence and properties of outflows may depend on the morphology and/or environment of the galaxy. 
However, major mergers alone are not sufficient to account for the entire AGN population \citep{Draper2012}, as they appear to be associated mainly with ultra-luminous infrared galaxies (ULIRGs) \citep{Treister2012,Glikman2015,Barrows2017,Barrows2023}. That is, only the most luminous AGN phases are associated with major mergers, while less luminous AGNs appear to be driven by secular processes \citep{Treister2012}.

Our goal is to test whether the host galaxy of AGNs behaves differently in the case with and without outflow signatures. Therefore, in this work, we will discuss and compare the properties of galaxies hosting mid-IR selected AGNs with and without outflow signatures.
This paper is structured as follows: Section \ref{sec:parent} summarizes the selection criteria for the sample used in this work and how we select and measure outflows. In section \ref{sec:host_properties} we describe some of the AGN and host galaxy properties such as \oiiil~luminosity, X-ray emission, stellar-mass, star formation rate (SFR), rest-frame colors and morphological properties.  We discuss our results in section \ref{sec:discussion}, and in section \ref{sec:conclusions} we summarize our results.

Throughout this work we will use the AB magnitude system \citep{oke} and we will assume a $\Lambda$CDM cosmology with H$_{0}$=70 km s$^{-1}$Mpc$^{-1}$, $\Omega_{M}$=0.3, $\Omega_{\Lambda}$=0.7.

\section{Sample selection}
\label{sec:parent}
\subsection{Parent Sample}
\label{sec:parent_sample}
Due to the diversity of the AGN population, no single AGN selection technique is complete. A specific technique may correctly identify one population of AGNs while missing others. For example, a selection based solely on X-ray or optical emission may overlook AGNs that are obscured by interstellar gas or dust. In contrast, MIR selection methods are more effective at detecting this population because they are less affected by extinction \citep[e.g.,][]{L04,S05,donley12,Dai2014}. Thus we have chosen to compare the host galaxies of AGNs selected in the IR to avoid other selection biases.
We derived a sample of mid-IR (MIR) selected and optically confirmed AGNs from the work of \cite{bornan2022}. Briefly, in that study, they investigated the properties of host galaxies of AGNs selected based on near-IR emission and MIR criteria, which were subsequently confirmed spectroscopically. The different selection techniques have varying completeness and reliability at selecting AGN \citep[see e.g.,][]{lacy2020}. 
In this work, we will focus on the AGN sample selected using the criteria of \cite{L04}, which presents a high level of completeness \citep{messias2014}, and confirmed through optical spectroscopy using the mass-excitation (MEx) diagram \citep{J11,J14} in order to obtain more confidence in the sample.
\cite{bornan2022} chose the AGN sample from the COSMOS field \citep{scoville} using the COSMOS2015 catalog \footnote{The catalog can be downloaded from \url{ftp://ftp.iap.fr/pub/from\_users/hjmcc/COSMOS2015/}} \citep{laigle} and the zCOSMOS redshift survey \citep{lilly07,lilly09}. The COSMOS2015 catalog provides photometric data across multiple wavelengths, ranging from UV to mid-IR, as well as estimates of stellar mass obtained using \lephare~\citep{Arnouts2002,Ilbert2006} following the method described in \cite{ilbert13}. For their analysis, \cite{bornan2022} utilized the Spitzer IRAC and MIPS bands: 3.6, 4.5, 5.8, 8.0, 24, and 70µm \citep{sanders07,lefloch09,ashby13,ashby15}.
The authors selected sources within the spectroscopic redshift range of 0.3$\leq$$z_{\rm sp}$$\leq$0.9 from the zCOSMOS DR3-bright catalog, ensuring reliable redshift estimates. This catalog provides a confidence parameter that ranges from insecure and probable redshift (Class 1 and 2, respectively), one broad AGN redshift (Class 18), one line redshift (Class 9), and secure and very secure redshift (Class 3 and 4, respectively). 
Each confident parameter is also assigned a confidence decimal, which is derived from 
repeat observations and by the consistency or otherwise with photometric redshifts. The confidence decimal ranges from .1 (spectroscopic and photometric redshifts are not consistent at the level of 0.04(1+z)), .3 (special case for Class 18 and 9, consistent with
photo-z only after the redshift is changed to the alternate redshift), .4 (no photometric redshift available) to .5, (spectroscopic redshift consistent within 0.04(1+z) of the photometric redshift). In the work of \cite{bornan2022} they selected all sources with classes 3.x and 4.x, where x can take the values 1, 3, 4, and 5. Using the photometric data, they performed a preselection of sources that satisfied the \cite{L04} color-color criteria, employing the bands [4.5]-[8.0] vs [3.6]-[5.8].
\cite{L04} identified the positions of QSOs detected in the Sloan Digital Sky Survey (SDSS) on a log(S8.0/S4.5) vs. log(S5.8/S3.6) color-color diagram and proposed the following relation to select AGN samples based on their MIR colors:

\begin{eqnarray}
\rm log(S_{5.8}/S_{3.6})>0.1, \nonumber\\
\rm log(S_{8.0}/S_{4.5}) > -0.2 \wedge \nonumber\\
\rm log(S_{8.0}/S_{4.5}) \leq  0.8\times \rm log(S_{5.8}/S_{3.6})+0.5
\end{eqnarray}

Here, $\wedge$ represents the logical AND operator (refer to Figure 1 from \citealt{bornan2022}). A total of 490 AGN candidates with good redshift estimates fall within the Lacy wedge and constitute the parent sample for this study.

As noted by several authors \citep[e.g.,][]{sajina2005,L07,donley12,barrows2021}, AGN IR-selection techniques can suffer from contamination by galaxies classified as star-forming (SF) or quiescent. Hence, it is advisable to apply a second selection criterion to ensure that the chosen sample predominantly consists of AGNs. Optical emission-line ratios can be used to diagnose the ionization mechanism, for example, using the BPT diagram \citep{bpt,kewley2001,kauffmann2003}. However, within the redshift range of the parent sample, only the H$\beta$ and \oiiil~ lines fall within the wavelength coverage of the zCOSMOS spectra. 
As an alternative to the BPT diagram, the MEx diagram utilizes the host galaxy mass and the quotient [O\,{\footnotesize III}] $\lambda$5007/\rm H$\beta$\, making it useful for selecting AGNs at intermediate redshifts. To measure the signal-to-noise ratio (S/N) of these emission lines, a single Gaussian profile fitting was performed on the H$\beta$ and \oiii~ lines. Only spectra with both H$\beta$ and \oiii~ S/N ratios greater than 3 were considered. 

After applying the IR selection criteria of \cite{L04}, 490 AGN candidates were selected. The selection criterion according to the MEx diagram yields a number of 122 AGNs and the restrictions on S/N on both \oiiil~and \hb~spectral lines, resulted in a final number of AGNs of 103.

We then select possible outflow candidates following the methods described in Sec \ref{sec:outflow_selection}.

\subsection{Outflow Candidates}
\subsubsection{Emission line measurements}
\label{line_measures}
We are interested in measuring \oiiil~as a tracer of possible outflows in our galaxy sample. In AGNs, this line is produced in the low-density narrow-line region, and any asymmetries are often attributed to outflowing (or inflowing) gas, with some parts of this wind obscured by surrounding dust. To measure the emission lines, we followed a two-step procedure.

\begin{figure}
\centering
\includegraphics[width=\columnwidth]{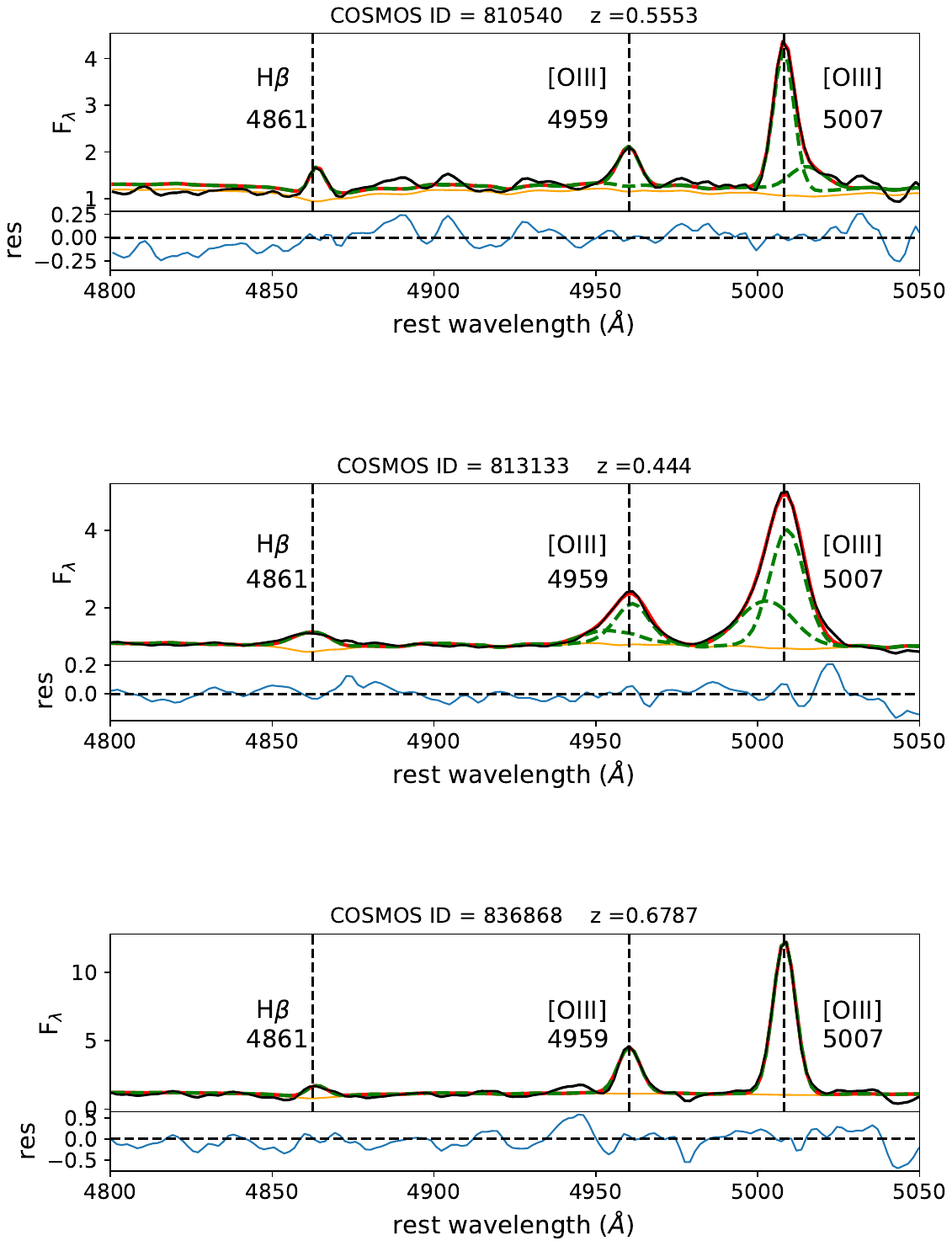}
\caption{Fit examples of different Gaussian configuration. \textit{Top panel:} The galaxy COSMOSID=810540 is one of the few objects in our sample with a red wing asymmetry in its [OIII]$\lambda$5007 line. \textit{Middle panel:} The object COSMOSID=813133 shows one of the strongest blue asymmetry in our sample. It is to be noted that for most of our sample with detected asymmetries in [OIII]$\lambda$5007 the asymmetric component was not significantly detected in [OIII]$\lambda$4959, nevertheless we will focus only on the former. \textit{Bottom panel:}  COSMOSID=836868 shows the [OIII] emission line well fitted with a single Gaussian. The flux density is in units of 10$^{-17}$ erg s$^{-1}$ cm$^{-2}$ \AA{}$^{-1}$ in all panels.}
\label{fig:spectra_examples}
\end{figure}

First, we shifted the spectra to the rest frame using the spectroscopic redshift values tabulated in the zCOSMOS catalog \citep{lilly07} and fitted the continuum using the Penalized Pixel-Fitting code (\ppxf\, \citealt{capp2004,capp2017}). Since the power-law component and Fe II emission are produced in the innermost regions of the nuclei, in type II AGNs, this contamination will be enshrouded by the surrounding obscuring dust. Therefore, we chose not to include these components in the continuum fitting \citep[see e.g.,][]{Oio2019}. Next, the emission lines were fitted using the {\sc ifscube} package\footnote{https://github.com/danielrd6/ifscube/} \citep{ifscube}. {\sc ifscube} is a Python-based software package primarily developed to perform analysis tasks in data cubes of integral field spectroscopy but has great flexibility to work with single-slit or fiber spectra. This package subtracts the stellar continuum obtained in the previous step and performs a constrained multiple-component fitting on the pure emission line spectrum.
In the first iteration, Oxygen lines (\oiii$\lambda$4959,5007) and H$\beta$ were fitted with a single Gaussian component. The velocity from the Gaussian fit of \oiiil~was assumed as the systemic velocity, and this value was reintroduced in \ppxf\ to bring each spectra in our sample to a frame of reference based on the observed wavelength of the peak of their \oiii~line.
Then, a second fit to the emission lines was performed, this time allowing up to two Gaussian components for the oxygen lines, while a single Gaussian component was sufficient for H$\beta$ in all cases. We present example fits for different cases of emission line profiles in Figure \ref{fig:spectra_examples}. We show two prominent cases of \oiii~ asymmetric emission, one with a red wing and one with a blue wing, as well as a typical example of a spectrum with a symmetrical line profile (i.e., single Gaussian).

To analyze the kinematics of the emission lines we employed a non-parametric scheme similar to the one utilized by \cite{robleto2021} and described in detail in \cite{Harrison2014}. In this approach, we solely consider the cumulative function of the synthetic line profile. Like \cite{robleto2021}, we do not attribute any physical significance to the individual Gaussian profiles.
The non-parametric velocity definitions employed in this study are the line width at 80\% of the flux (W$_{80}$) and the velocity offset ($\Delta v$), which is measured from the velocities corresponding to the 5th and 95th percentiles of the overall emission-line profiles. Specifically, $\Delta v$ is calculated as $\Delta v$ = ($v_{05}$ + $v_{95}$)/2. We chose to work with these values because W$_{80}$ and $\Delta v$ are more sensitive to asymmetries in the base of the line profile compared to, for instance, the full width at half maximum (FWHM). Both W$_{80}$ and $\Delta v$ were measured from the fitted line profile of \oiiil. 

\subsubsection{Line constrains: outflow selection}
\label{sec:outflow_selection}
As mentioned in section \ref{line_measures}, for our selection criteria we fitted the \oiii~ lines with up to two Gaussian components, namely a ``core" component and a broader ``asymmetric" component. As summarized in section \ref{sec:intro},  there are several somewhat arbitrary ways to select optical emission line outflows.
To decide whether a second component was necessary or not we imposed the following criteria:

\begin{itemize}
    \item The flux of both Gaussian components must be at least 1$\sigma$ greater than the noise measured in the vicinity region of \oiiil~ (5030 $-$ 5080 \AA{}).
    \item The velocity difference between the two components should be greater than the sum of their respective velocity errors.
    \item The line width W$_{80}$ must be greater than 300 km s$^{-1}$.
\end{itemize}

That is we require both Gaussian profiles to be detectable over the noise of the spectra, and for the asymmetric component to be distinguishable from the core component in terms of velocity shift. The third criteria is introduced to ensure that we are in the presence of a broad asymmetrical profile, with the imposed velocity cut being approximately twice the instrumental velocity dispersion.
Lastly, we required that the flux of both the \oiiil~and \ocl~lines to be at least 1$\sigma$ above the noise of the vicinity of the line.
By applying these criteria, we retained a total of 91 AGNs, 26\% (24) of which exhibited an asymmetric profile. Out of the 24 AGNs, 21 displayed a blue asymmetry, while 3 objects showed a red wing ($\Delta$v $>$ 0 km s$^{-1}$), indicating possible outflows and inflows, respectively. Although we decided not to fit a possible Fe II emission component, we took special care with objects exhibiting a red asymmetry. For these three objects, we re-fitted the continuum by including an Fe II template, but in no case was this component necessary according to our fitting model.  Figure \ref{fig:appendix1} in the Appendix \ref{appendix} presents the spectra for all the objects with asymmetrical profile. All spectra are in rest frame considering the peak of the \oiiil~ profile.

If we apply stricter selection criteria, for instance, if we require that each Gaussian component's flux of \oiiil~be greater than 2$\sigma$ and 3$\sigma$, we are left with 18 and 15 asymmetric profiles, respectively. The different choice of S/N requirement in the Gaussian profile does not significantly change our sample or the following conclusions. At 3$\sigma$ it is dropped from the asymmetry sample the galaxy COSMOS ID = 810540, which is the object with the highest red asymmetry ($\Delta$v = 118 km s$^{-1}$, W$_{80}$ = 573 km s$^{-1}$). The most significant change in our sample arises from disregarding the velocity difference between the two components. If we were to disregard this criterion, the final sample of potential outflows would consist of 46 objects (instead of the 24 selected), representing 50\% of the total.

We estimated the uncertainties for the fitted line parameters by performing 1000 Monte Carlo iterations for each spectrum. The errors for each Gaussian parameter were then determined as the 1$\sigma$ dispersion obtained from the 1000 Monte Carlo runs.

\begin{figure}
    \centering
    \includegraphics[width=\columnwidth]{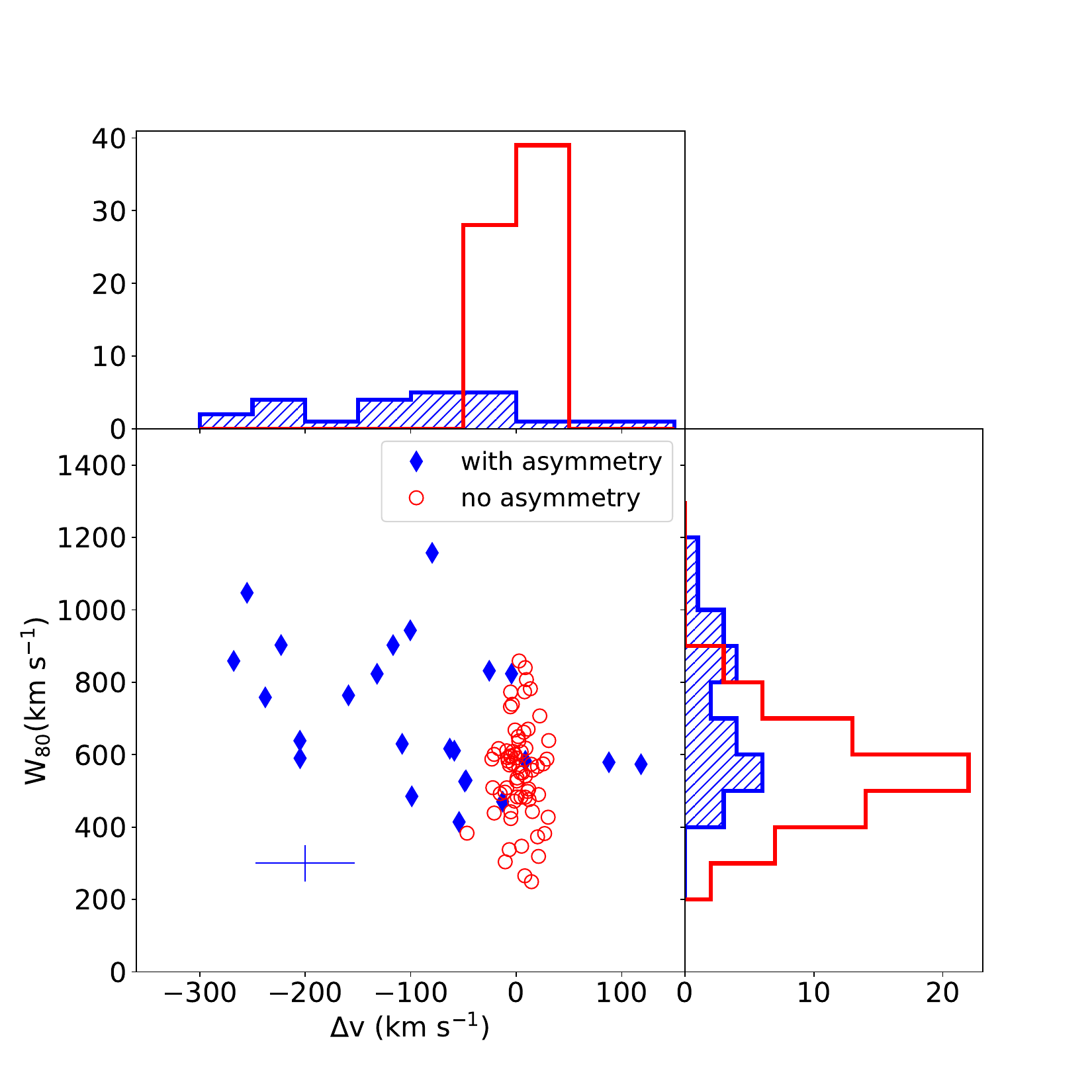}
    \caption{Relation between the velocity difference measured at the 5th and 95th percentiles ($\Delta$v) and the full line profile width at 80\% of the flux (W$_{80}$). Diamond symbols indicate galaxies with asymmetrical profiles and open circles represent the absence of an asymmetrical profile in [OIII] line, on the bottom left is shown the median error for galaxies with asymmetrical profiles. \textit{Top panel:} $\Delta$v distribution in units of km s$^{-1}$, shaded histogram corresponds to galaxies with asymmetrical profiles, while blank histogram corresponds to galaxies with symmetrical profiles. \textit{Right panel:} distribution of W$_{80}$ in units of km s$^{-1}$.}
    \label{fig:w80_deltav}
\end{figure}

In Figure \ref{fig:w80_deltav}, we present the relation between W$_{80}$ and $\Delta v$. The majority of objects with asymmetry exhibit negative values of velocity difference, indicating potential outflows. The distribution of velocity dispersion and velocity differences shows a wide range of values for objects with asymmetric profiles, while it is narrower for objects with no discernible asymmetries. In the following we will characterize the distributions by their median values and adopt the error as the standard deviation from the mean. The distribution of $\Delta v$, shown in the top panel of Figure \ref{fig:w80_deltav}, has a median value of $-$89 $\pm$ 100\,km\,s$^{-1}$ for AGNs with asymmetry, and a median value of 2 $\pm$ 14\,km\,s$^{-1}$ for galaxies with no asymmetry. The median value of W$_{80}$ is 634 $\pm$ 190\,km\,s$^{-1}$ for galaxies with asymmetric profiles, and a median of 567 $\pm$ 132\,km\,s$^{-1}$ for objects with no asymmetry.

Our estimated outflow velocities align with those found in previous studies. For example, \cite{matzko2022}, who examined a large sample of galaxy pairs involving AGNs and SF galaxies, used W$_{80}$ as a measure of outflow velocity, taking it as 1.088$\times$FWHM. They reported a mean value of approximately 700\,km\,s$^{-1}$ for all AGN subsamples. \cite{zakamska2014} measured \ocl~and \oiiil~lines in a sample of obscured QSOs and found a median value for W$_{80}$ of 752\,km\,s$^{-1}$. From a sample of 16 type 2 AGNs observed with Integral Field Unit (IFU), \cite{Harrison2014} found W$_{80}$ values ranging between 600 and 1500\,km\,s$^{-1}$.
In the case of highly luminous type II QSOs (log(L$_{\rm \oiii}$)$>$42.5), the velocity widths are considerably higher, with W$_{80}$ ranging from approximately 1000 to 5500\,km\,s$^{-1}$  \citep{Zakamska2016,Storchi-Bergmann2018}. As depicted in Figure \ref{fig:w80_deltav}, there is a tendency for galaxies with larger blue asymmetries to exhibit higher velocity dispersions for the ionized gas. Henceforth, we disregard galaxies with red asymmetries and refer to AGNs with outflows as those objects with $\Delta$v $<$ 0 km s$^{-1}$.

\section{AGN and Host Galaxy Properties}
\label{sec:host_properties}

\subsection{\oiiil~luminosity}
\begin{figure}
\centering
\includegraphics[width=\columnwidth]{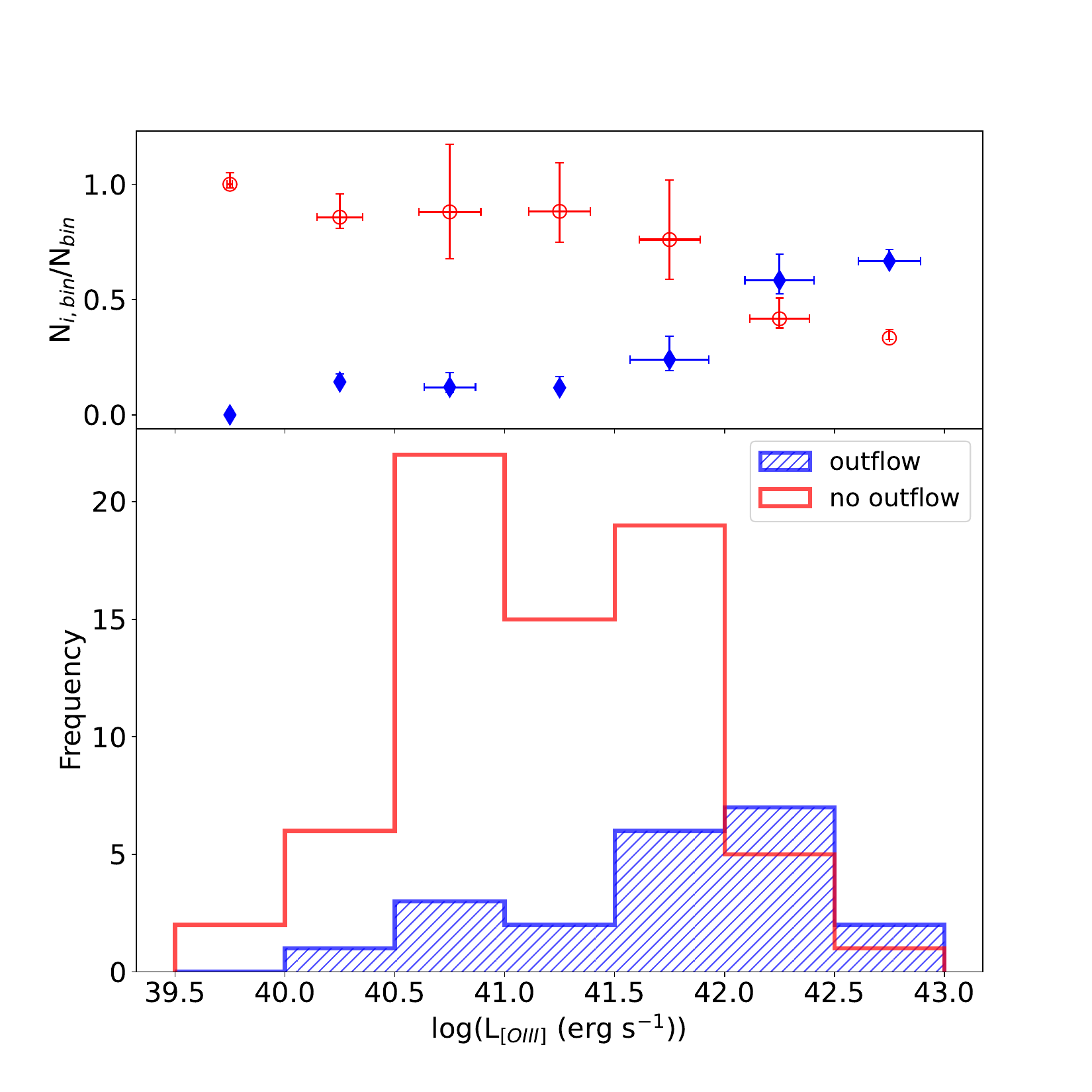}
\caption{Distribution of the \oiiil~luminosity. \textit{Top panel:} Fraction of AGNs with (closed diamonds) and without (open circles) asymmetric profiles as a function of luminosity. Vertical error bars correspond to the 68.3\% binomial confidence interval for a beta distribution \citep{Cameron2011}. The position in the x-axis corresponds to the middle value in each bin and their associated error is the 1$\sigma$ standard deviation.}
\label{fig:hist_LOIII}
\end{figure}

We calculate the AGN luminosity from [OIII] to test if the AGNs with and without outflow features have intrinsically different luminosities. The luminosity for the \oiiil~emission line was computed with the standard formula:
\begin{equation}
    L_{\rm \oiii} = \frac{4\pi d_{L}^{2}}{(1+z)}f_{\rm \oiii}
\end{equation}
where d$_{L}$ is the luminosity distance and f$_{\rm \oiii}$ is the \oiiil~line flux. As in \cite{bornan2022} we have not corrected the \oiii~line flux for extinction. The distribution of the \oiiil~luminosity (L\oiii) for the sample of 91 galaxies is shown in Figure \ref{fig:hist_LOIII}. There is a tendency for galaxies with asymmetric profiles towards higher \oiiil~luminosity values, presenting a median value of log(L\oiii)=41.9 (erg s$^{-1}$) against a median value of log(L\oiii)=41.1 (erg s$^{-1}$) for the L\oiii~ of galaxies with no asymmetric profile. The greater fraction of AGNs with asymmetrical profiles as a function of L\oiii~is evident in the top panel of Figure \ref{fig:hist_LOIII} where we show the incidence for each type of galaxy per [OIII]$\lambda$5007 luminosity bin.
In the Figure, uncertainties on the incidence fraction are given by the binomial beta distribution quantile technique with a 1$\sigma$ (68.3\%) confidence interval \citep{Cameron2011}, the error in luminosity value for each bin are given by the standard deviation of the mean.

%In Figure \ref{fig:deltav_LOIII} we show the relation between the velocity difference $\Delta$v and the outflow velocity W$_{80}$ as a function of \oiii~luminosity. Although with a high dispersion, we can see a trend between $\Delta$v and L\oiii~for objects with outflows. We find a Spearman correlation coefficient S$_{p}$ = 0.36 with a p\_value = 0.08.
Some authors find a good and even strong correlation between W$_{80}$ and L\oiii~\citep[e.g.,][]{perna2017,wylezalek2020,scholtz2021,matzko2022}, however we found no significant correlation between these parameters. Considering the relation between W$_{80}$ and L\oiii~and without making any difference on AGNs with and without signs of an outflow we find a Spearman correlation coefficient S$_{p}$ = 0.1 with a p\_value = 0.3, if we consider the sample with asymmetrical profiles S$_{p}$ = 0.01 with p\_value = 0.9, and for the objects with symmetrical profiles S$_{p}$ = $-$0.08 and p\_value = 0.5.

\subsection{X-ray properties}
We investigate whether there is a relationship between the incidence of AGN with and without outflows and X-ray luminosity and hardness ratio.

The COSMOS2015 catalog \citep{laigle} also provides information about X-ray photometry. In their catalog they include fluxes and fluxes errors from the previous \textit{Chandra} COSMOS Survey (C-COSMOS) \citep{Elvis2009,puccetti2009,civano2012}, X-ray detected sources from XMM-COSMOS \citep{hasinger07,cappeluti07,brusa2010}, the matches with the \textit{Nu}STAR Extragalactic Survey \citep{civano2015} and also with the \textit{Chandra Cosmos-Legacy} Survey \citep{civano,marchesi}.
The latter catalog is a 4.6 Ms Chandra program over 2.2 deg$^{2}$ of the COSMOS field, containing 4016 X-ray sources down to a flux limit off 2$\times$10$^{-16}$erg s$^{-1}$cm$^{-2}$ in the 0.5$-$2 keV band. 
For our sample of 91 AGNs we found 34 matches with the XMM-COSMOS catalog, 29 matches with the C-COSMOS, 5 matches with \textit{Nu}STAR COSMOS survey and 42 with the \textit{Chandra COSMOS-Legacy} Survey. The catalog of \cite{civano} encompasses all of the previous X-ray matches so we decided to use the fluxes and hardness ratio provided by this catalog.

\begin{figure}
\centering
\includegraphics[width=\columnwidth]{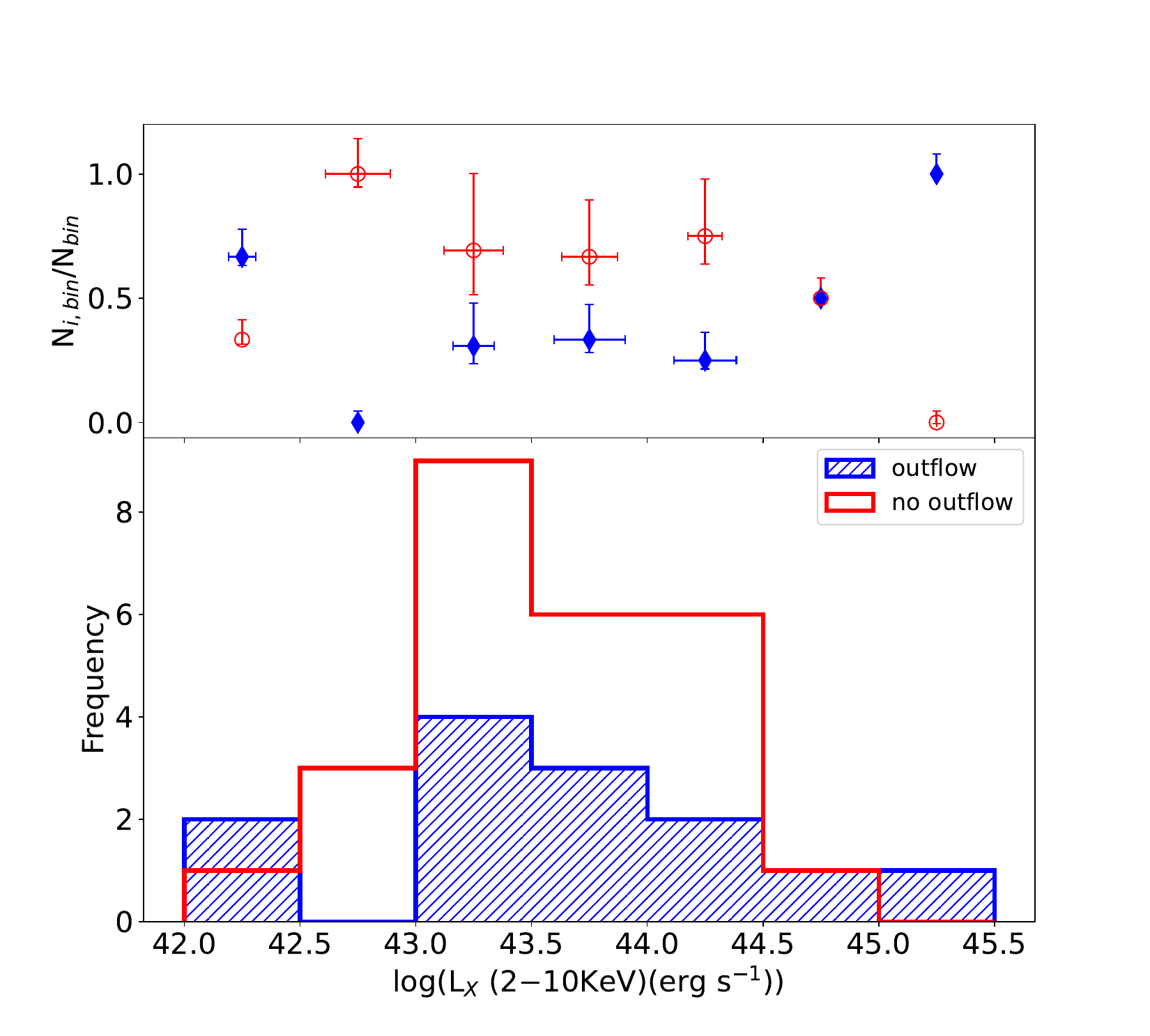}
\caption{Distribution of the X-ray luminosity in the 5$-$10 KeV band. In the top panel is shown the incidence of each class of line profile per luminosity bin. Errors were estimated as in Figure \ref{fig:hist_LOIII}.}
\label{fig:hist_LX}
\end{figure}

In Figure \ref{fig:hist_LX}, we present the distribution of the X-ray luminosity for the AGN selected galaxies with and without outflow signatures calculated as:
\begin{equation}
    L_X = 4\pi d^{2}_{L}f_{x}(1+z)^{\Gamma - 2}~{\rm erg~s^{-1}}
\end{equation}

Here, d$_L$ is the luminosity distance in cm, f$_x$ is the X-ray flux in units of erg s$^{-1}$cm$^{-2}$ in the hard-band and the photon index was assumed to be $\Gamma$ = 1.8 \citep{Tozzi06}.
In the top panel of Figure \ref{fig:hist_LX} we plot the fraction of AGNs with (closed diamonds) and without (open circles) asymmetric profiles as a function of the X-ray luminosity. We can see that the incidence is practically constant except for the bins of the highest L$_X$ luminosity where we see an increase in the fraction of AGNs with outflow. This result is in agreement with the work of \cite{Harrison2016} where studying a sample of 89 AGN at z $\gtrsim$ 0.6, they find a larger fraction of AGNs with high wind velocities (W$_{80}$ $>$ 600 km s$^{-1}$), for the objects with L$_{X}$ $>$ 10$^{43}$ (erg s$^{-1}$).
Later, \cite{perna2017} confirmed these results in a larger sample of X-ray AGNs ($\sim$ 500 SDSS/X-ray objects), with outflow signatures in their optical spectra. They also found a higher fraction of AGNs with outflows at the highest X-ray luminosity bins.
As in the case with \oiii~luminosity, we found no significant correlation between the kinematic parameters and the X-ray luminosity.

\begin{figure}
\centering
\includegraphics[width=\columnwidth]{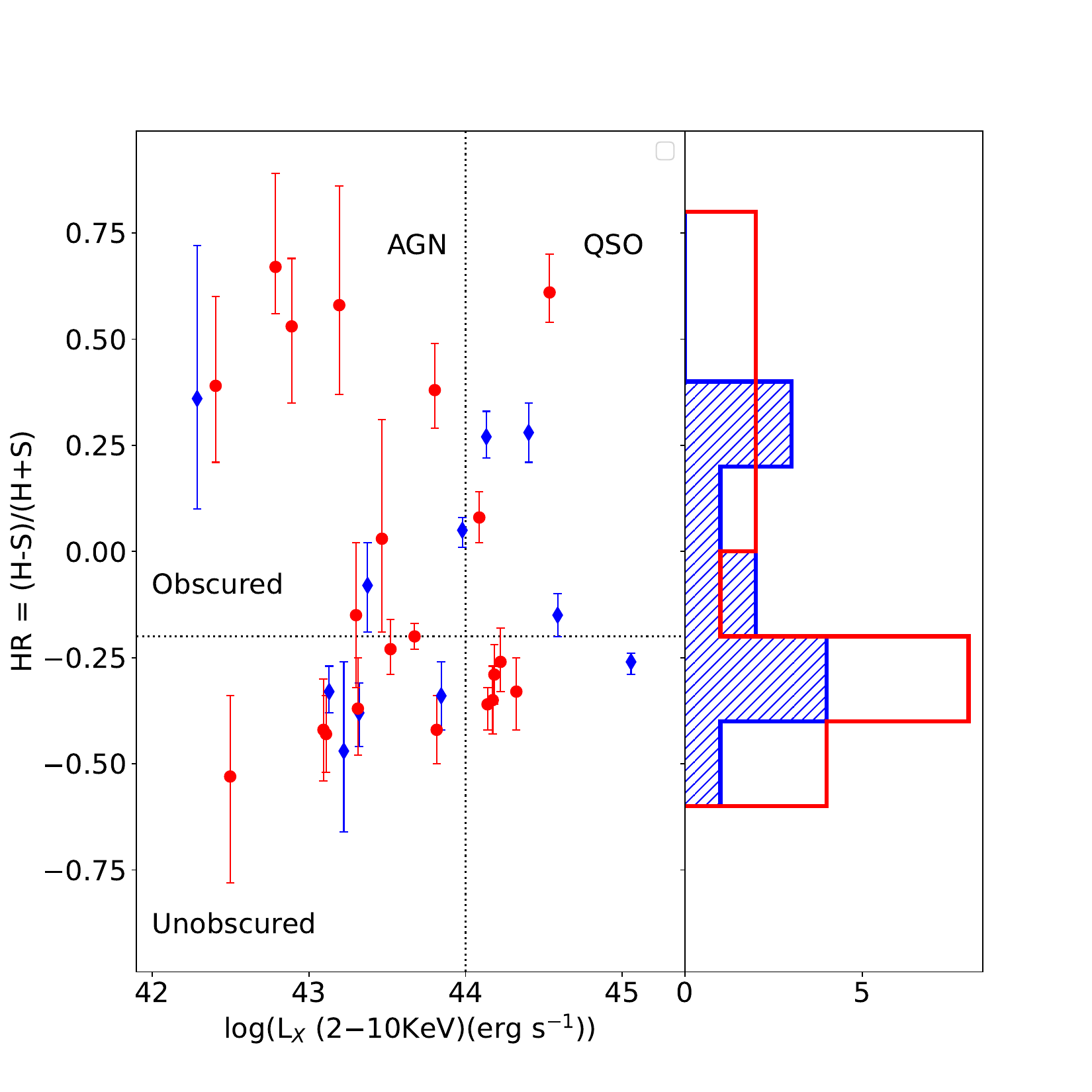}
\caption{Hardness Ratio as a function of the X-ray luminosity in the 2$-$10 KeV band. Errors in HR correspond to the 90\% confidence limits from \cite{civano2015}.}
\label{fig:HR_LX}
\end{figure}

We also want to compare whether there is a preference for AGN with outflows to be found in X-ray obscured (i.e., high neutral hydrogen column densities or high hardness ratio values) with the AGN without optical outflow signatures.
The hardness ratio (HR) is defined as:
\begin{equation}
HR = \frac{H - S}{H + S},      
\end{equation}
where S refers to the net count rate in the soft band 0.5$-$2 keV and H is the count rate in the hard band 2$-$7 keV. So stated, HR provides an indication about the flatness of an X-ray spectrum. 
In Figure \ref{fig:HR_LX} we plot the HR as a function of the 2$-$10 keV luminosity for the objects in our sample with X-ray emission. Several authors made use of the HR to set apart obscured from unobscured sources in the X-rays at all redshifts \citep[e.g.,][]{gilli09,treister09,marchesi}. This is due to the the fact that soft X-ray emission of obscured AGNs tend to be absorbed, while hard X-ray are able to escape. 
We have taken as limiting value HR=$-$0.2 (dashed horizontal line in the plot) which correspond for a source with a neutral hydrogen column density, N$_H$ $>$ 10$^{21.6}$ cm$^{-2}$ \citep{gilli09}.

The vertical dashed line indicates a typical limit used in the X-rays to separate AGNs from QSOs \citep{treister09}. On the right panel we show the distribution of HR for objects with detected outflows (blue) and without outflows (red).
As it can be seen in the figure, the distribution of HR is skewed towards softer spectra. Taking into account the low number of X-ray detection in AGNs with outflow, their distribution seems to be bi-modal with a mean value for HR=$-$0.09$\pm$0.28 with a slightly higher number of obscured sources (55\%). Inversely, 43\% of AGNs without outflows have HR$\geq$-0.2 (i.e., obscured). While the obscured sources with outflows seem to be evenly distributed in the areas corresponding to AGN and QSO, there seems to be a preference for the unobscured sources with optical outflows to be located in the AGN region, being 37\% of them found in this area.
In contrast, we can also see in that figure that in general the most luminous sources with outflows show a harder x-ray spectra than the most luminous sources without outflow.

\subsection{Stellar mass}

\begin{figure*}
\centering
\includegraphics[width=\columnwidth]{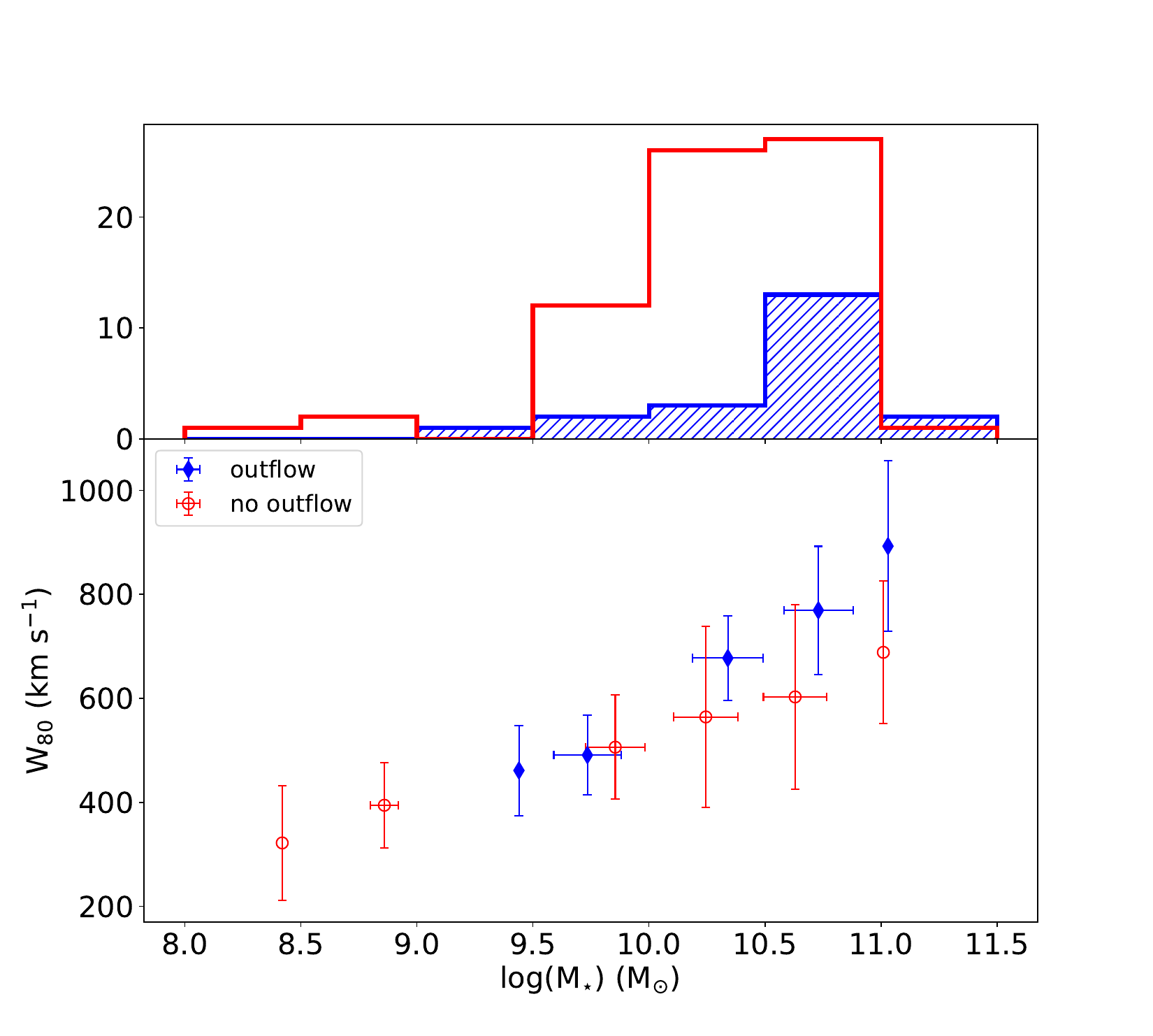}
\includegraphics[width=\columnwidth]{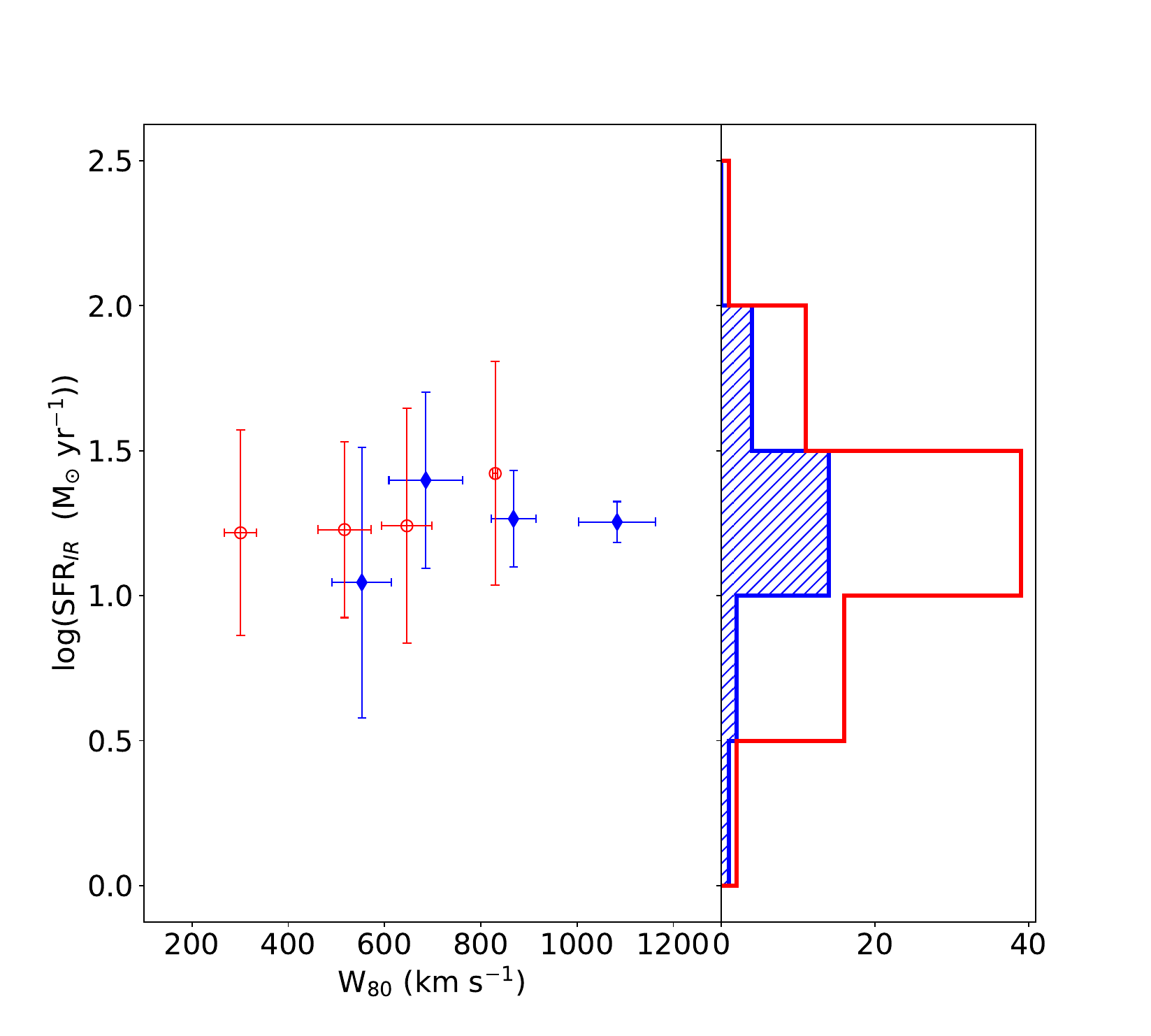}
\caption{Left: Relation between the [OIII] velocity (W$_{80}$) and the stellar mass. Right: Star formation obtained from FIR+mm SED fitting as function of the gas velocity. We can see that higher mass galaxies present the higher [OIII] velocities, while the overall SFR does not seems to be affected by the presence of an outflow.}
\label{fig:W80_mstar}
\end{figure*}

Feedback is thought to be the main mechanism behind the co-evolution of the AGN host galaxy and its supermassive black hole. This would be responsible for regulating star formation and galaxy growth in AGNs.
One of the main methodologies to determine the stellar-mass and the SFR of galaxies is through modeling the ultraviolet (UV) to infrared (IR) spectral energy distributions (SEDs) of galaxies. 
Because the beam sizes of FIR/(sub)mm detectors are very large, the fluxes from individual galaxies are sometimes difficult to measure if they are close to other sources. This introduces heavy source confusion (blending) which makes it difficult to correctly measure the SED and determine the SFR. \citet{Liu2018} developed a new method called ``super-deblending" approach for obtaining prior-fitting multiband photometry for FIR/(sub)mm data sets in the GOODS-North field. 
For this reason, in this work we use stellar mass and SFR determinations obtained in the Super-deblended catalog in the COSMOS fields by \citet{Jin2018}. In the work of \cite{Jin2018} they applied this method to 194428 galaxies in the COSMOS field, covering data from  Spitzer,  Herschel, SCUBA2, AzTEC, MAMBO and NSF's Karl G. Jansky VLA at 3 GHz and 1.4 GHz. They use SED fitting techniques following the same approach as the one presented in \cite{Liu2018}; namely they include four SED  components in the fitting procedure: 1) a stellar component \citep{BruzualCharlot2003} with a Small Magellanic Cloud attenuation law; 2) a mid-infrared AGN torus component  \citep{mullaney2011}; 3) a dust continuum emission from the \cite{magdis2012} library; 4) a power-law radio continuum with an evolving $_{qIR}$=2.35$\times$(1+z)$^{-0.12}$+log(1.91) \citep{Delhaize2017,Magnelli2015}.
In this section we make use of their compiled values for the stellar mass \citep{laigle,Muzzin2013} and SFR$_{IR}$ computed from the integrated 8$-$1000$\mu$m infrared luminosities resulting from the FIR+mm SED fitting, assuming a Chabrier IMF \citep{Chabrier2003}, and excluding any AGN contamination as derived from the SED fitting.
Figure \ref{fig:W80_mstar} shows the median values of outflow velocity in bins of stellar mass on the left panel, and the median star formation rate as a function of W$_{80}$ on the right. We observe a clear trend of higher velocities in \oiii~line toward higher galaxy masses with the outflows being detected in galaxies with stellar masses higher than log(M$_{star}$ (M$_{\odot}$))$=$9.4. 

\begin{figure}
\centering
\includegraphics[width=\columnwidth]{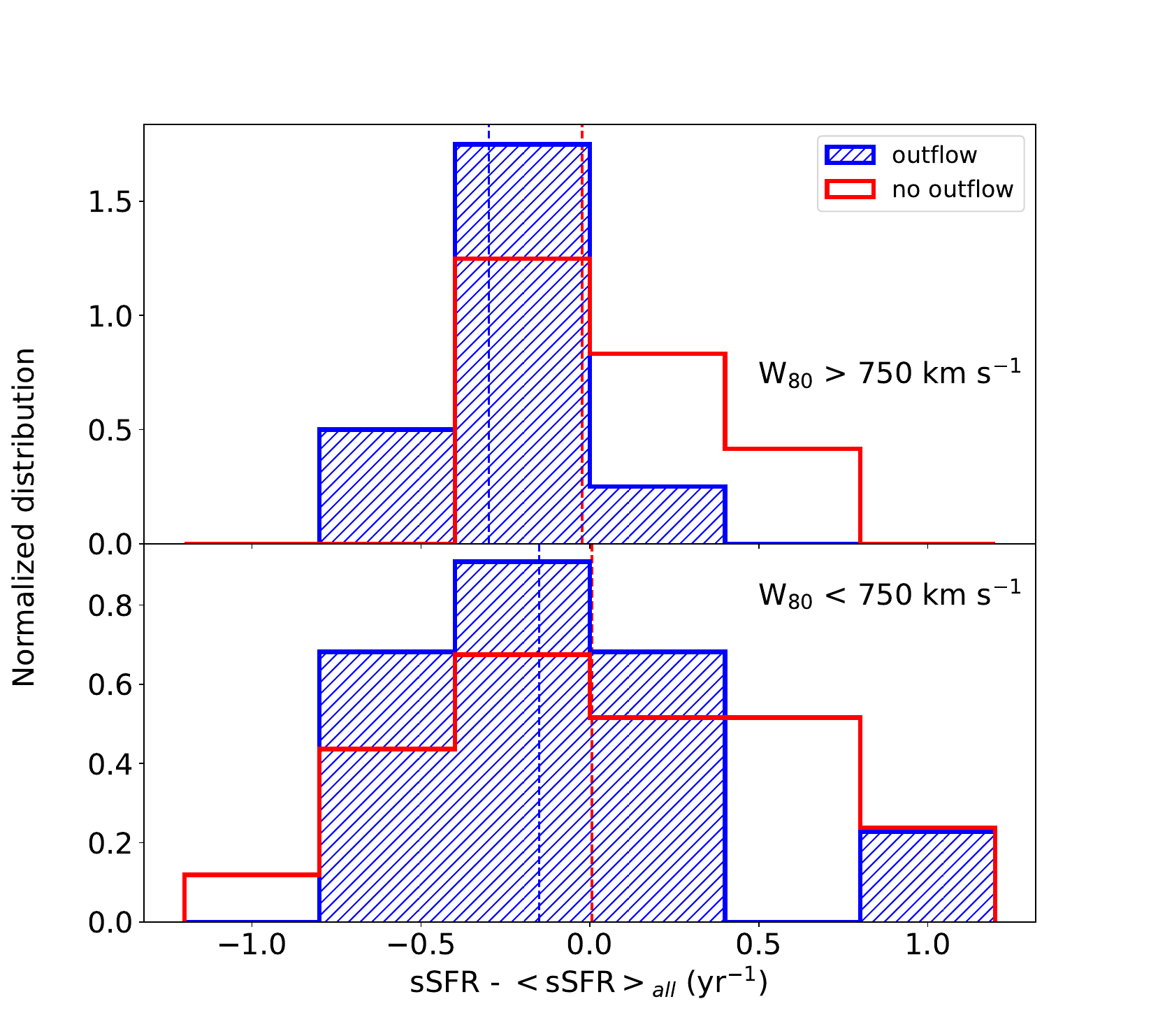}
\caption{Distribution of the relative sSFR depending on outflow velocity of AGNs with outflow (blue) and without outflow (red), with respect to the mean value of the whole sample. The median value of each distribution is denoted with vertical dashed lines. High velocity outflow AGNs tend to show lower sSFR than AGNs with symmetrical line profiles.}
\label{fig:dist_deltasSFR}
\end{figure}

AGN driven outflows are often invoked to explain the quenching of star formation by ejecting the interstellar medium (ISM) gas and preventing the cooling and infall of intracluster medium (ICM) gas on larger scales \citep[e.g.,][]{dimatteo2005,Croton2006,Choi2018,Cresci2015,kmos3d2019,Piana2022}. 
We find a median star formation rate of log(SFR$_{IR}$ (M$_{\odot}$ yr$^{-1}$))= 1.3 $\pm$ 0.3 for AGNs with outflows, and log(SFR$_{IR}$ (M$_{\odot}$ yr$^{-1}$))= 1.2 $\pm$ 0.3, for AGNs without outflows. We see on the right panel of Figure \ref{fig:W80_mstar} that there is no impact on the star formation rate throughout the velocity range studied. 
To continue investigating any possible difference in the star formation among AGNs with and without outflows, we now turn our attention to the specific star formation rate (sSFR=$\frac{SFR}{M_*}$). We compare the distribution of the relative sSFR in comparison to the mean sSFR of the whole sample in Figure \ref{fig:dist_deltasSFR}. We have divided the sample taking as limiting value W$_{80}$ $=$ 750 km s$^{-1}$ which is the median W$_{80}$ for AGNs with outflow. We can see that in both cases the relative sSFR for AGNs with outflow is lower than for AGNs without outflow. The bigger difference appears to be at higher velocities where their median values of difference from the mean sSFR are -0.3 $\pm$ 0.2 and $-$0.02 $\pm$ 0.3 respectively. This result, together with the mass scaling relation, could hint towards a feedback impact in the most massive galaxies. However, this difference is not significant enough as to claim a quenching in the star formation due to the outflow.
Our results agree with those by \cite{Ward2022}, who studied three cosmological simulations with AGN feedback and found that AGNs are preferentially found in galaxies with high gas fractions and sSFR. According to this, the outflows observed in AGNs do not necessarily imply the quenching of star formation, even if this negative feedback occurs over long timescales.

\subsection{Quiescent and star forming galaxies}
\begin{figure}
\centering
\includegraphics[width=\columnwidth]{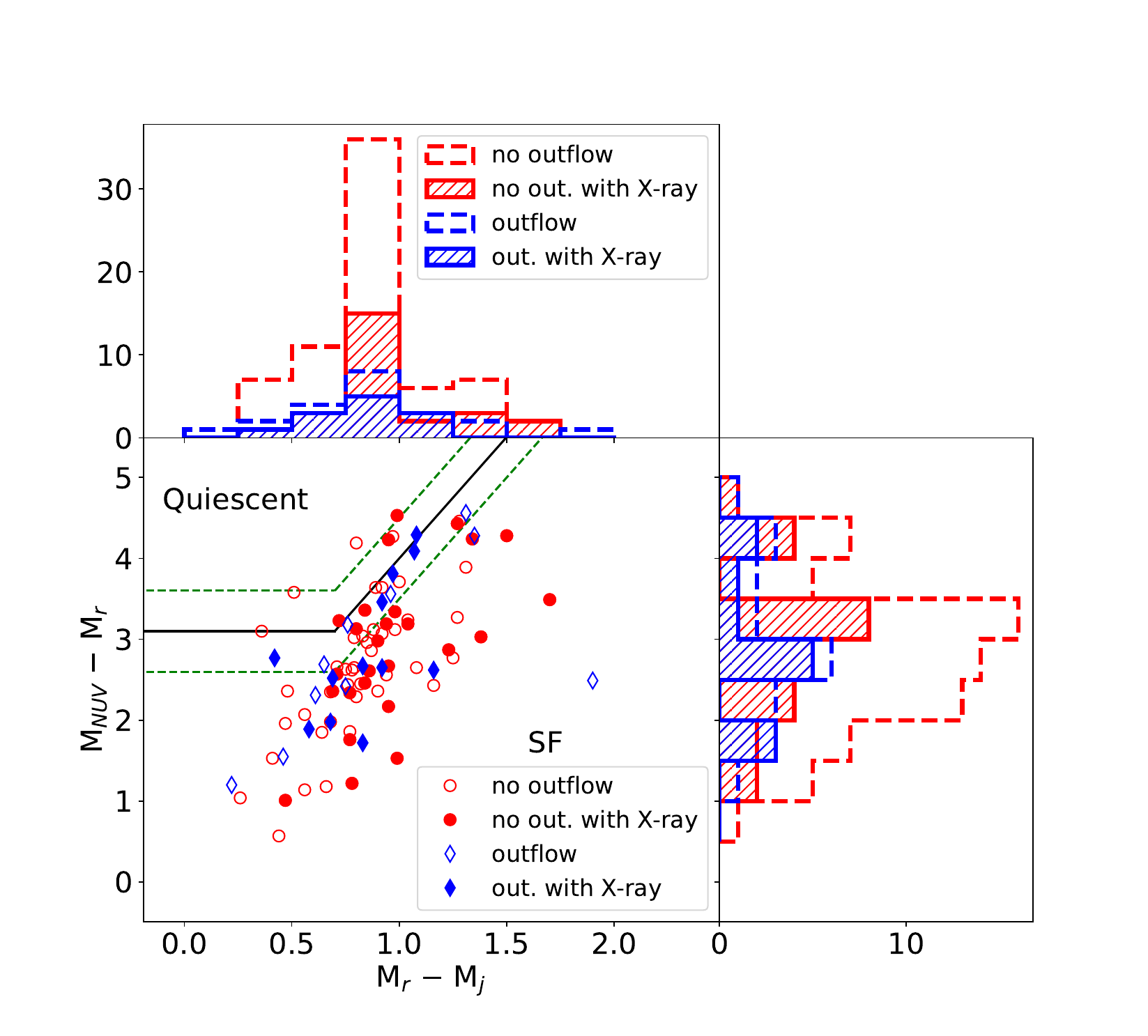}
\caption{Rest-frame (M$_{NUV}$ $-$ M$_r$) vs. (M$_r$ $-$ M$_J$) color-color diagram for AGN host galaxies with outflows (blue) and without outflow signatures (red). Solid black line mark regions which separates quiescent (upper-left corner) and star-forming galaxies \citep{ilbert13}. Dashed green lines encompass the transitional region Green Valley. Filled circles represent AGNs with detected X-ray emission. Color distributions are included in the upper and right panels for the objects with and without outflows (dashed lines, blue and red respectively) and for objects with X-ray emission (dashed histograms).} 
\label{fig:dust_SF}
\end{figure}

With the help of color-color diagrams, we can study whether AGNs with and without outflows are found in galaxies with different stellar populations.
Rest-frame color-color diagrams have been largely used to separate populations of quiescent and star-forming galaxies \citep[e.g.,][]{chester64,hogg2003,bell2004,baldry2006,arnouts2007,williams09,arnouts13,ilbert13,bouquin2015,moutard2016,moutard2016b,pacifici2016,Foltz2018,xu2020}. \cite{williams09} introduced a selection method using dust corrected optical colors, and considering the optical and near-IR emission of galaxies at redshifts z $\leq$ 2, the rest-frame U$-$V versus V$-$J color$-$color (hereafter UVJ) diagram. The rest-frame NUV $-$ r$^{+}$ versus r$^{+}$ $-$ J diagram introduced by \cite{ilbert2010}, was presented as an alternative to the UVJ diagram of \cite{williams09}, as NUV $-$ r$^{+}$ is more sensitive to the history of star formation activity \citep[e.g.,][]{Martin2007,arnouts2007}, while the r-band is more sensitive to the amount of stellar mass, formed over the course of a galaxy's history.
Figure \ref{fig:dust_SF} shows the rest-frame M$_{NUV}$ $-$ M$_{r}$ vs M$_{r}$ $-$ M$_{J}$ color-color diagrams for the AGNs with and without asymmetric component (blue and red symbols, respectively). The solid line separate regions occupied by quiescent and star-forming galaxies as defined by \cite{ilbert13}, where galaxies with M$_{NUV}$ $-$ M$_r$ $>$ 3(M$_r$$-$M$_J$)+1 and M$_{NUV}$ $-$ M$_r$ $>$ 3.1 are considered as quiescent. The blending between dusty star-forming galaxies and quiescent galaxies is avoided with this selection given that dust absorption would shift star-forming galaxies along a diagonal axes from the bottom left to the top right of Fig. \ref{fig:dust_SF}. With the increased fraction of red galaxies observed across time \citep{Faber2007}, it has been postulated an evolutionary path for galaxy populations from blue to red. In the color-magnitude diagram there is a region in between the blue and red galaxy populations, called Green Valley (GV) which would be inhabited by a transition population called Green Valley galaxies \citep{wyder07,salim07,salim14}. The boundaries used to define this region vary among authors, with the limiting values being e.g., 4$<M_{NUV}-M_{r}<$4.5 \citep{Martin2007}, 3.5$<M_{NUV}-M_{r}<$4.5 \citep{salim2009}, 3.2$<M_{NUV}-M_{r}<$5 \citep{mendez2011} or by a linear relation considering other bands as in \cite{McNab2021} with 2(M$_{V}-M_{J}$)$+$1.1 $\leq$ (M$_{NUV} - M_{V}$) $\leq$ 2($M_{V} - M_{J}$) $+$ 1.6. In this work as in \cite{davidzon2017}, we consider a $\pm$ 0.5 mag. shift from the limit defined by \cite{ilbert13}, shown in Fig. \ref{fig:dust_SF} with dashed green lines. In top and right panels we also include the corresponding color distributions for each sample.
We find that 59\% of the AGNs with outflows and 74\% of AGNs with symmetrical \oiii~profiles are located in the star-forming region, while 41\% of the galaxies with outflows and 24\% without outflows reside in the green valley region, and only two objects can be found in the quiescent zone (both without outflow). As it can be seen, we find an excess of 15\% for galaxies with star-formation and no evidence of outflows in their optical spectra. On the other hand, galaxies with outflows are 17\% more likely to be found in the GV than AGN with no outflows. Regardless, the color distributions of AGNs with and without outflows are indistinguishable. For the M$_r$ $-$ M$_j$ colors for AGNs with outflows and without outflows we obtain a median value of 0.8 $\pm$ 0.2 and 0.8 $\pm$ 0.3, while for the M$_{NUV}$ $-$ M$_r$ colors we get a median of 2.6 $\pm$ 0.7 and 2.6 $\pm$ 0.9 respectively.
In Fig. \ref{fig:dust_SF} we also show the objects with detected X-ray emission as filled points, and dashed histograms. For the AGNs with no outflows and X-ray emission we find the following percentages: (SF, GV, quiescent)$=$(77, 19, 4)\%. For the sample of AGNs with outflows we find (SF, GV, quiescent)$=$(58, 42, 0)\%. When constraining our sample to objects with X-ray emission we obtain more than double AGNs with outflow in the green valley region, with respect to AGN without outflow, while the large majority of AGN without outflows and X-ray emission can be found in the SF region.

\subsection{Morphology}
\subsubsection{S\'ersic profiles}
In order to address whether the occurrence of outflows is related to their host galaxy morphology, we perform three independent analysis derived from observer-frame optical data. 
First, we study the host galaxy morphological properties as given by the S\'ersic index, making use of the data provided by the Advanced Camera for Surveys General Catalog (ACS-GC) \citep{gri12}. This catalog used publicly available data obtained with the Advanced Camera for Surveys (ACS) instrument on the Hubble Space Telescope to construct a photometric and morphological database. The imaging data used to construct the ACS-GC was collected from four surveys covering 469,501 galaxies: the All-wavelength Extended Groth Strip International Survey \citep[AEGIS,][]{davis2007}, the Great Observatories Origins Deep Survey \citep[GOODS,][]{GOODS1,GOODS2}, the Cosmological Evolutionary Survey \citep[COSMOS,][]{scoville}, and the Galaxy Evolution from Morphologies and SEDs \citep[GEMs,][]{gems1,gems2}. The data used in this work correspond to the COSMOS survey which covers an area of 1.8 deg$^{2}$ in the F814W filter, with a limiting AB magnitude of 26.0. At the mean redshift of our sample ($z$ = 0.59) this broadband filter images covers the spectral region corresponding to H$\beta$ + \oiii$\lambda\lambda$4959,5007, which is closest to the \oiii~feature we used to find the outflow features. 
In \cite{gri12} they employed an automated fitting method called {\small GALAPAGOS} \citep{galapagos,GEMS} to measure structural parameters such as the S\'ersic index \citep{sersic1968}. With a search radius of 5 arcsec, we find a total of 100 objects from our initial sample (of 103 objects) present in this catalog. 

As in \cite{bornan2022}, we also exclude objects with $n$ = 0.2 and $n$ = 8 that are likely to correspond to erroneous fits or systematics, therefore we constrain the S\'ersic index to be in the range 0.2 $<$ $n$ $<$ 8. With this restriction and after applying the line signal cut described in Section \ref{sec:outflow_selection}, we are left with 75 objects.
We separate the galaxies into two main classes: (1) late type galaxies, spirals or disc dominated with 0.2 $<$ $n$ $<$ 2.5; (2) early-types or elliptical galaxies, with 2.5 $<$ $n$ $<$ 8 \citep{gri12}.
According to their S\'ersic index values and without making any distinction regarding their \oiii~profiles, we find a percentage of 45 $\pm$ 5\% disc dominated or late type galaxies and 55 $\pm$ 5\% corresponding to early type galaxies. 
While for the 13 galaxies with \oiii~asymmetrical profiles and valid S\'ersic indexes we obtain that 23$_{-7}^{+14}$\% of them correspond to late-types/spiral galaxies, 77$_{-14}^{+7}$\% early-types or elliptical galaxies. Among the 62 galaxies with symmetrical line profiles we find an even distribution of late type or disc dominated galaxies (50 $\pm$ 6\%) and early types or bulge dominated galaxies (50 $\pm$ 6\%). 
We estimated the uncertainties for the computed fractions (percentages) according to a Bayesian approach using a 68.3\% (1$\sigma$) confidence interval, as explained in \cite{Cameron2011}.

\subsubsection{Non-parametric morphology}
Continuing with our morphological analysis, we use in second place quantitative measures of the distribution of light. By doing so, we can avoid a-priory assumptions about the distribution of the light. Here we use three morphological parameters that are commonly used in non-parametric methods for galaxy classification, the Asymmetry  index (A) \citep{abraham1996,conselice2000}, the Gini coefficient (G) \citep{abraham2003} and the moment of the brightest 20\% of galaxy flux (M20) \citep{lotz2004,lotz2008}. 
The Asymmetry index of a galaxy is determined by rotating the image by 180$^{\circ}$ and then subtracting it from the original image, and then adding up the absolute value of the differences in intensity at each pixel location. This total value is then compared to the original flux of the galaxy.
The Gini coefficient, which was initially proposed as an economic indicator to assess wealth distribution within a population, has also been employed in astrophysics to quantify the inequality in the distribution of light across pixels in a galaxy. A value of Gini=1 would imply that all the light is concentrated in a single pixel, whereas a Gini coefficient of 0 indicates that the light is uniformly distributed across all pixels. M20 is a measure of the concentration of light in a galaxy. It is defined as the second order moment of the brightest 20\% of a galaxy's pixels.  It is a useful tool to distinguish between normal galaxies and non-symmetric objects, and to identify galaxies that have recently undergone a merger.
We use the morphological parameters from the catalog presented by \cite{cassata2007}. This catalog provides information on non-parametric diagnostics of galaxy structure derived from images from the Hubble Space Telescope ACS, for 232022 galaxies up to a limiting magnitude \textit{I$_{AB}$} = 25.

\begin{figure}
\centering
\includegraphics[width=\columnwidth]{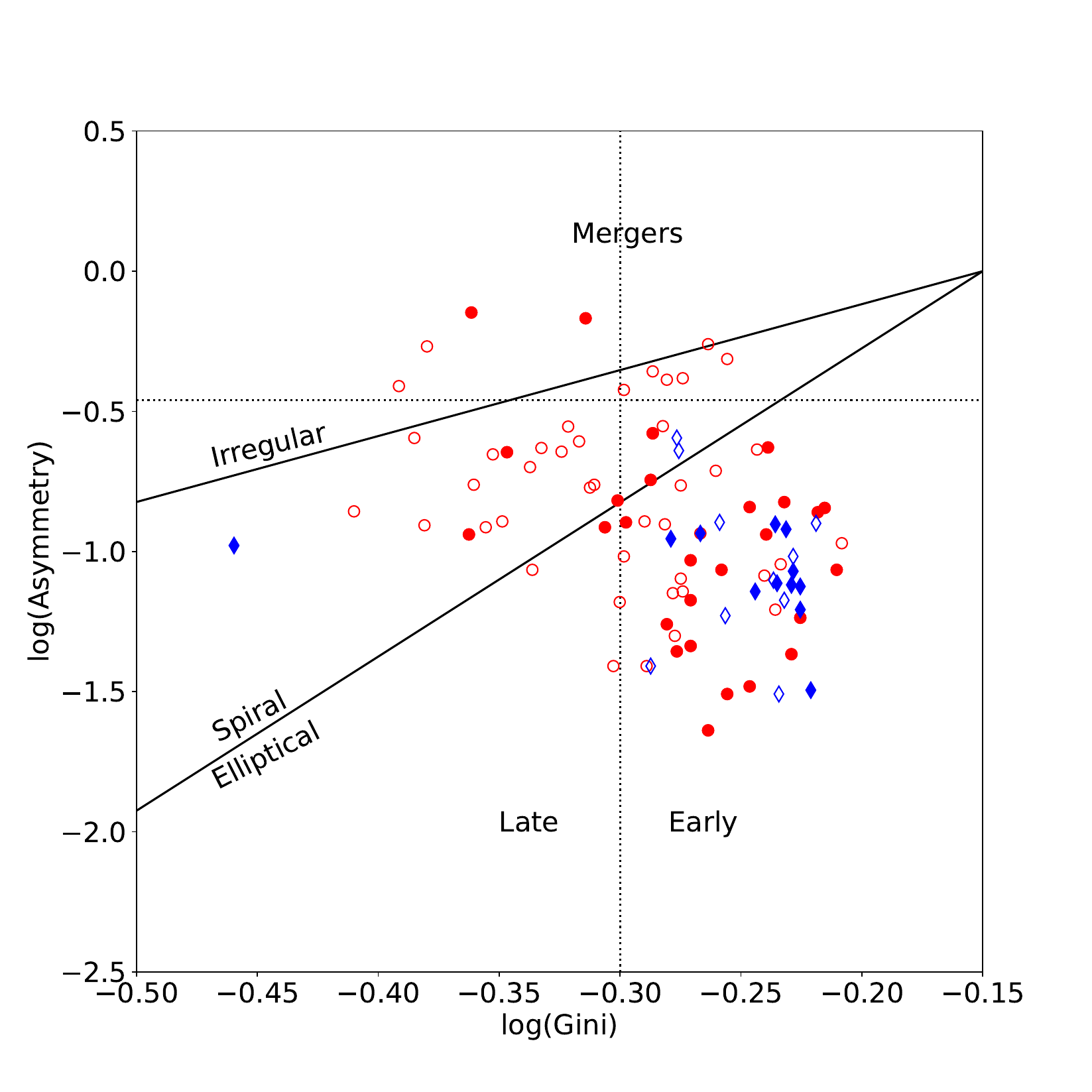}
\includegraphics[width=\columnwidth]{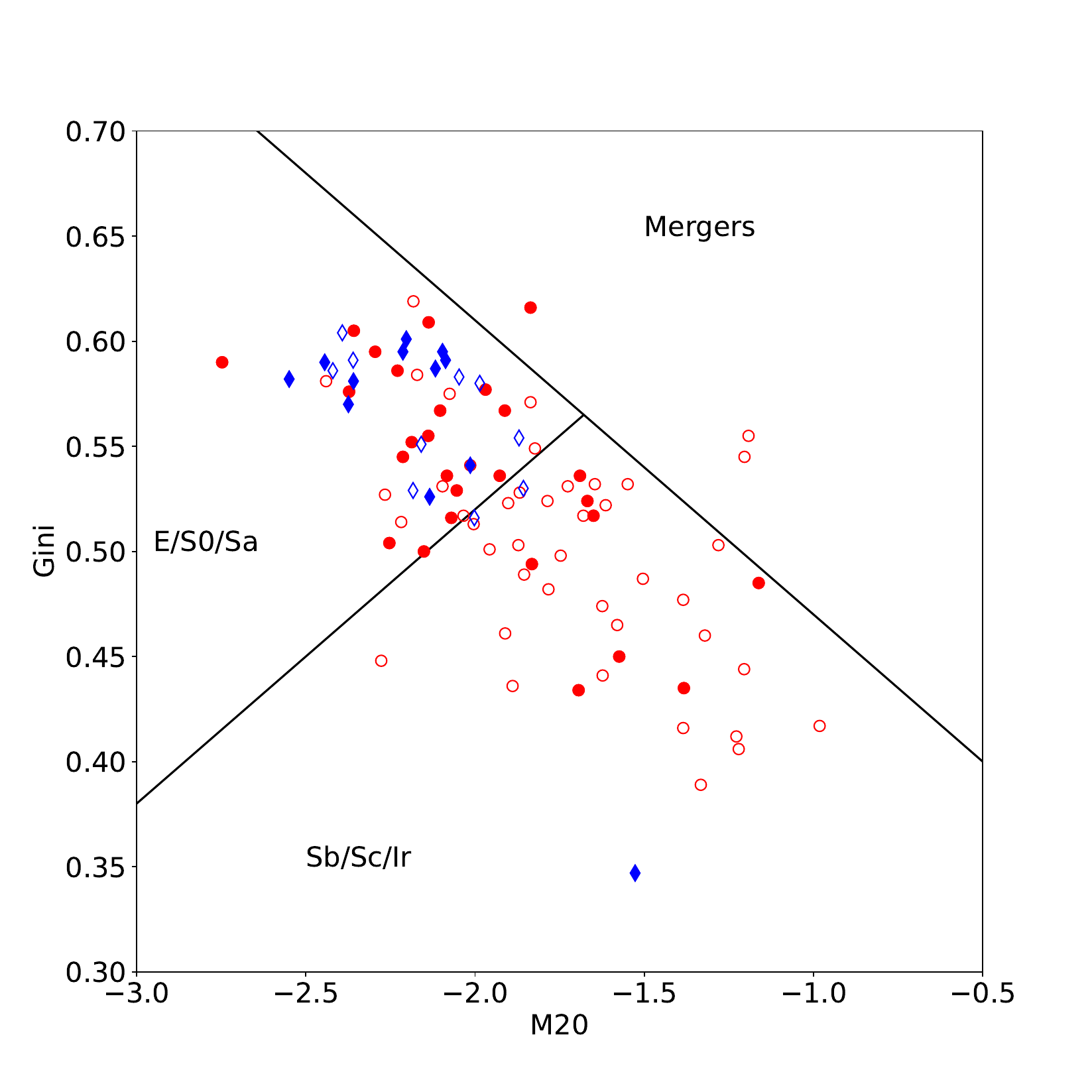}
\caption{Top: Asymmetry parameter versus Gini coefficient for AGNs with outflow signatures (blue) and symmetrical \oiii~profiles (red). The unfilled circles represent AGNs without X-ray detection. Solid lines shows regions of predominantly irregular, spiral, and elliptical types from \cite{capak07}. Dashed line at log(Asymmetry) = $-$0.46 \citep{conselice2003} shows the dividing line above which objects are expected to be major mergers. Vertical dotted line at log(Gini)= $-$0.3 separates late and early-type galaxies according to \cite{Abraham2007}.
Bottom: Gini coefficient vs M20, symbols are the same as in the top panel. Solid lines show regions of mergers, Sb/Sc/Irr and E/S0/Sa galaxy types taken from \cite{lotz2008}.}
\label{fig:non-par_morph}
\end{figure}

In Figure \ref{fig:non-par_morph} top panel, we plot Asymmetry versus Gini coefficient for the sample of AGNs with symmetrical \oiii~profiles (red circles) and for AGNs with outflows (blue diamonds). We include dividing lines between regions of predominantly irregulars, spirals, and elliptical morphological types taken from \cite{capak07}. The solid line between irregular and spiral galaxies is defined as log$_{10}$(Asymmetry)=2.353$\times$log$_{10}$(Gini)+0.353, while the solid line between spiral and elliptical galaxies is defined as log$_{10}$(Asymmetry)=5.5$\times$log$_{10}$(Gini)+0.825. We also include a horizontal dashed line to separate the region where mayor mergers are expected, defined by log(Asymmetry) $>$ $-$0.46 \citep{conselice2003} and a vertical dashed line at log(Gini)= $-$0.3 which separates late-type from early-type galaxies according to \cite{Abraham2007}.
In Fig. \ref{fig:non-par_morph}, bottom panel, we show the Gini parameter against M20. Here, we include dividing lines between the regions of mergers, Sb/Sc/Irr and E/S0/Sa galaxy types taken from \cite{lotz2008}, according to the following definitions, Mergers: G $>$ $-$0.14$\times$M20 + 0.33; Early(E/S0/Sa): G $\leq$ $-$0.14$\times$M20 + 0.33, and G $>$ 0.14$\times$M20 + 0.80; Late(Sb/Ir): G $\leq$ $-$0.14$\times$M20 + 0.33, and G $\leq$ 0.14$\times$M20 + 0.80. We note in both panels of Fig. \ref{fig:non-par_morph}, that AGNs with outflow signatures are preferentially located in the region corresponding to early-type/elliptical galaxies, with 95$_{-9}^{+1}$\% of them found in the area defined by \cite{Abraham2007} and 86$_{-11}^{+4}$\% inside the areas defined by \cite{capak07} and \cite{lotz2008}. While for the AGNs with no asymmetric component in \oiii, 55$_{-4}^{+7}$\% are early type galaxies according to \cite{Abraham2007} and \cite{capak07}, and 41$_{-5}^{+6}$\% according to the limits of \cite{lotz2008}. We do not obtain any galaxies with outflows in the region corresponding to mergers, and 14$_{-2}^{+5}$\%, 7$_{-1}^{+4}$\%, and 4$_{-1}^{+4}$\% of AGN without outflows are mergers in accordance to \cite{conselice2003}, \cite{capak07} and \cite{lotz2008} respectively.
%We find that 3.6 per cent of Type 1 AGN are located in the region occupied by irregular galaxies, 18.8 per cent are identified with spiral galaxies and 77.6 per cent with elliptical galaxies. While the majority (80 per cent) of Type 1 objects are located in the region occupied by elliptical or compact objects. Only 2 per cent are located in the region of spiral galaxies. In Fig. 6, right-hand panel, we find that 7.7 per cent of Type 1 AGNs are located in the region identified by mergers. This result is in agreement with those found by Chang et al. (2017), who studied a sample of 0.5 <z< 1.5 AGNs selected by their mid-IR power-law emission. These authors do not find a high merger rate in an obscured AGN sample.

\subsubsection{Visual Classification}
\begin{figure}
\centering
\includegraphics[width=\columnwidth]{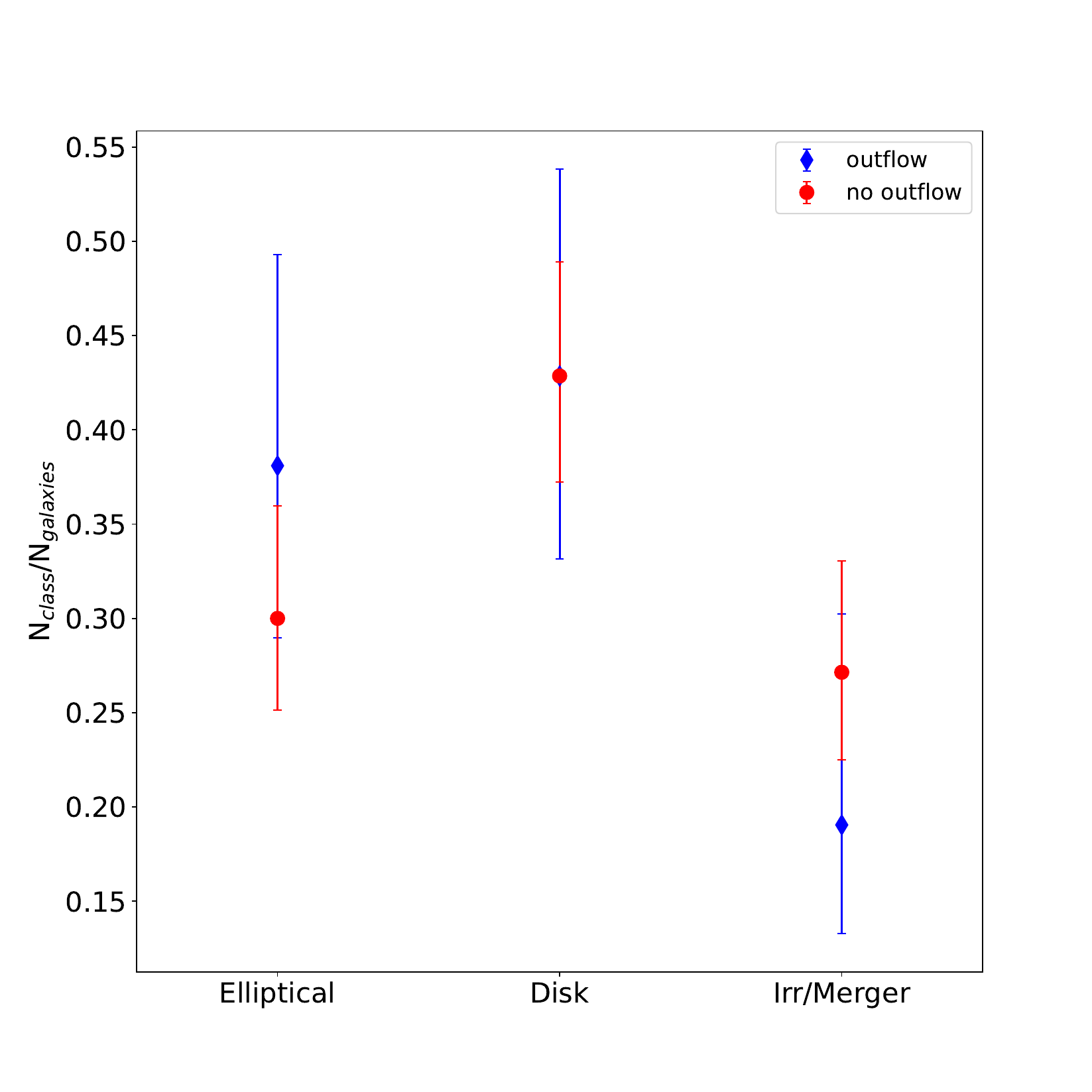}
\caption{Fraction of galaxies with (without) outflows in blue (red) for each morphological class from visual classification. The error bars correspond to the 68.3\% binomial confidence interval for a beta distribution \citep{Cameron2011}.}
\label{fig:frac_visual_morph}
\end{figure}

In this section we apply a visual classification criterion to estimate the morphology of the AGN host galaxies.
We classify the galaxies in our sample into three visual classes: elliptical (bulge dominated), spiral (disk dominated) and irregular/merger. This classification aims at describing the dominant morphology of the host galaxy. The objects were examined independently by the authors using the HST ACS F814W band images. To assign a galaxy to each category we requested a simple majority of votes. In Figure \ref{fig:frac_visual_morph}, we show the fraction of galaxies with and without outflows (blue and red symbols respectively) in each morphological type. We compute the error in each class according to a Bayesian approach using a 68.3\% (1$\sigma$) confidence interval, as explained in \cite{Cameron2011}. We obtain that AGNs with outflows are found preferentially in galaxies with elliptical (38$_{-9}^{+11}$\%) and disk (43$_{-9}^{+10}$\%) morphology with a slightly higher incidence than AGNs with no outflows in the elliptical class type. Conversely, the fraction of AGNs in merger (or irregular) galaxies with outflows (19$_{-5}^{+11}$\%) is lower than the AGNs in irregular host galaxies with no outflows (27$_{-4}^{+6}$\%).

By comparing with the non-parametric morphological classification, we find that the separation limit of \cite{capak07} recovers a 97\% of the elliptical galaxies, while the separation of log(Gini) $>$ $-$0.3 of \cite{Abraham2007} recovers 90\% of them. In the case of visually classified spiral (late types) galaxies, we see that they present a large scatter in their Gini index values, with $\sim$60\% of them found in the region for early-types. This result is in agreement with the works of \cite{Abraham2007} and \cite{kartaltepe2010}. A possible explanation is that spiral galaxies with a prominent bulge or significant AGN contribution will have a greater Gini concentration parameter and a lower M20 thus leading to a miss-classification.

\section{Discussion}
\label{sec:discussion}
\subsection{Outflow incidence}
\begin{figure}
\centering
\includegraphics[width=\columnwidth]{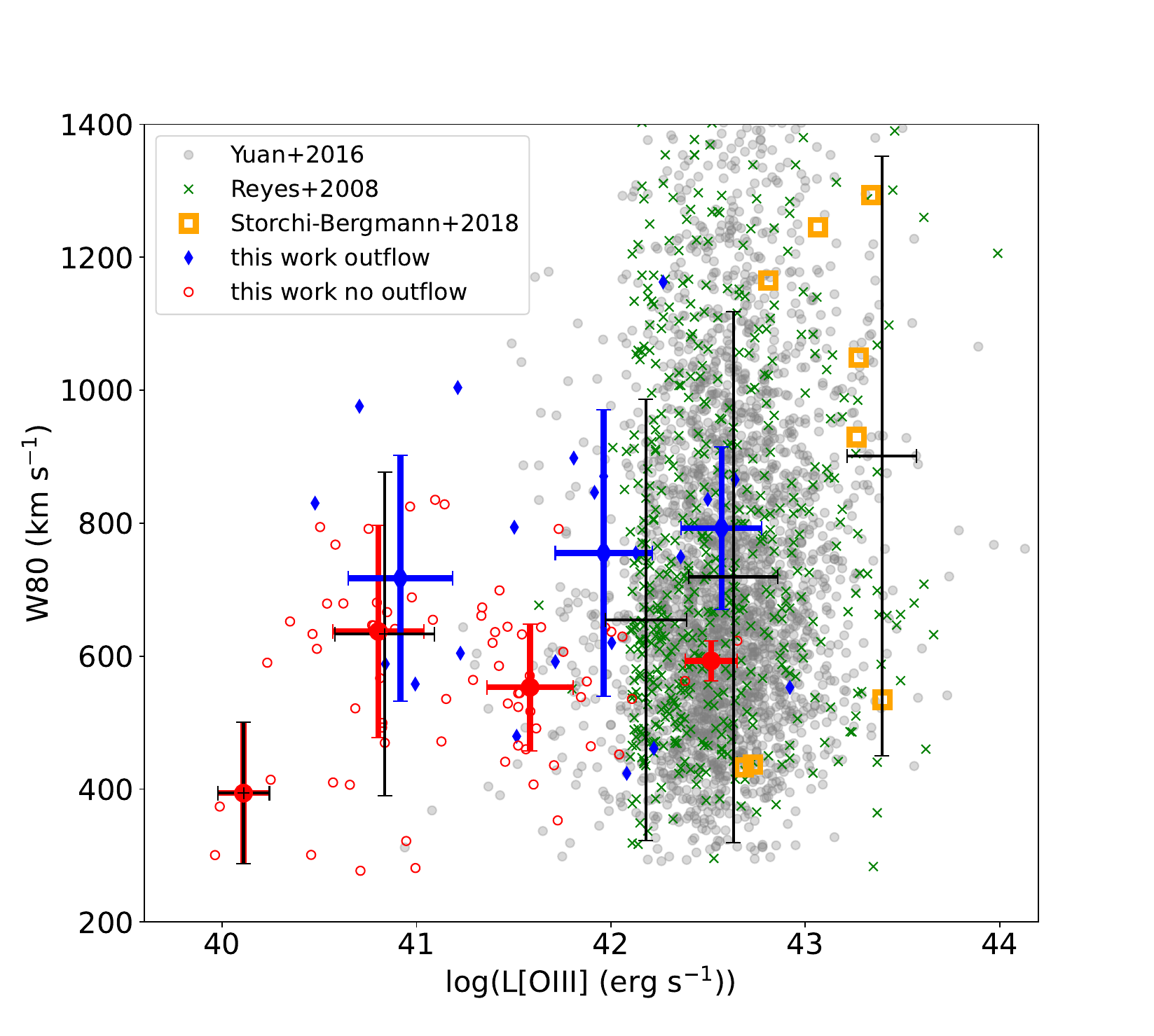}
\caption{W$_{80}$ as a function of \oiii~ luminosity. Blue filled diamonds and red open dots refer to the objects used in this work with and without outflows respectively; grey circles and green crosses shows type 2 Quasars from \cite{yuan2016} and \cite{reyes2008}; orange squares represent type 2 QSOs from \cite{Storchi-Bergmann2018}. Large symbols represent the median velocity in bins of \oiii~ luminosity for our sample (red and blue), and for the combination of all the objects in the four works (black). Error-bars correspond to the standard deviation in each bin. }
\label{fig:w80_LOIII_comp}
\end{figure}

%Analysing the \oiii line profiles
We can see from Figures \ref{fig:hist_LOIII} and \ref{fig:hist_LX} a higher incidence of AGNs containing outflow at increasing \oiii~and X-ray luminosity. This is consistent with previous works, where a higher fraction of galaxies with outflows are found at higher luminosities \citep[e.g.,][]{perna2017,avery2021,matzko2022}{}{}.
We would expect that the correlation reported between bolometric luminosity and outflow velocity \citep{Veilleux2013,spoon2013,fiore2017}, to be also present with the \oiiil~ luminosity. However, we do not find a clear trend between the outflow velocity and the total \oiii~ line luminosity nor with the X-ray luminosity. This could be partly explained because of the low \oiii~ line luminosity of our objects, while the trend is more noticeable at increased luminosity ranges. \cite{matzko2022} studied the merger impact on the outflow properties, derived from the \oiii~line profile, at a similar luminosity range than our sample. They report that a noticeable increase in the average outflow velocity is evident only in the most luminous \oiiil~ bin.
To have a broader perspective of the place our objects occupy within the context of the overall type 2 AGN population, in terms of their \oiii~ luminosities and velocities, we show in Figure \ref{fig:w80_LOIII_comp} a comparison with previous works. In this Figure, we compare our sources with the 2920 type 2 QSOs at a redshift range of 0.4 $\lesssim$ z $\lesssim$ 0.65, and \oiii~ luminosities in the range of log(L\oiii)$\sim $41$-$44 erg s$^{-1}$, from \cite{yuan2016} (grey symbols); the $\sim$560 luminous (log(L\oiii) $\gtrsim$41.6 erg s$^{-1}$) from \cite{reyes2008} (green crosses); and nine luminous (log(L\oiii) $>$ 42 erg s$^{-1}$) type 2 QSOs with redshifts 0.1 $<$ z $<$ 0.5 from \cite{Storchi-Bergmann2018} (orange squares). We compute the median values of velocity in luminosity bins for our objects with outflows (large blue symbols), and without outflows (large red symbols), and also combining all the data-sets plotted here (black symbols). We can see that, although with large dispersion, our objects seem to follow the overall W$_{80}$ $-$ luminosity trend.

\subsection{Compaction related to outflows}
\begin{figure}
\centering
\includegraphics[width=\columnwidth]{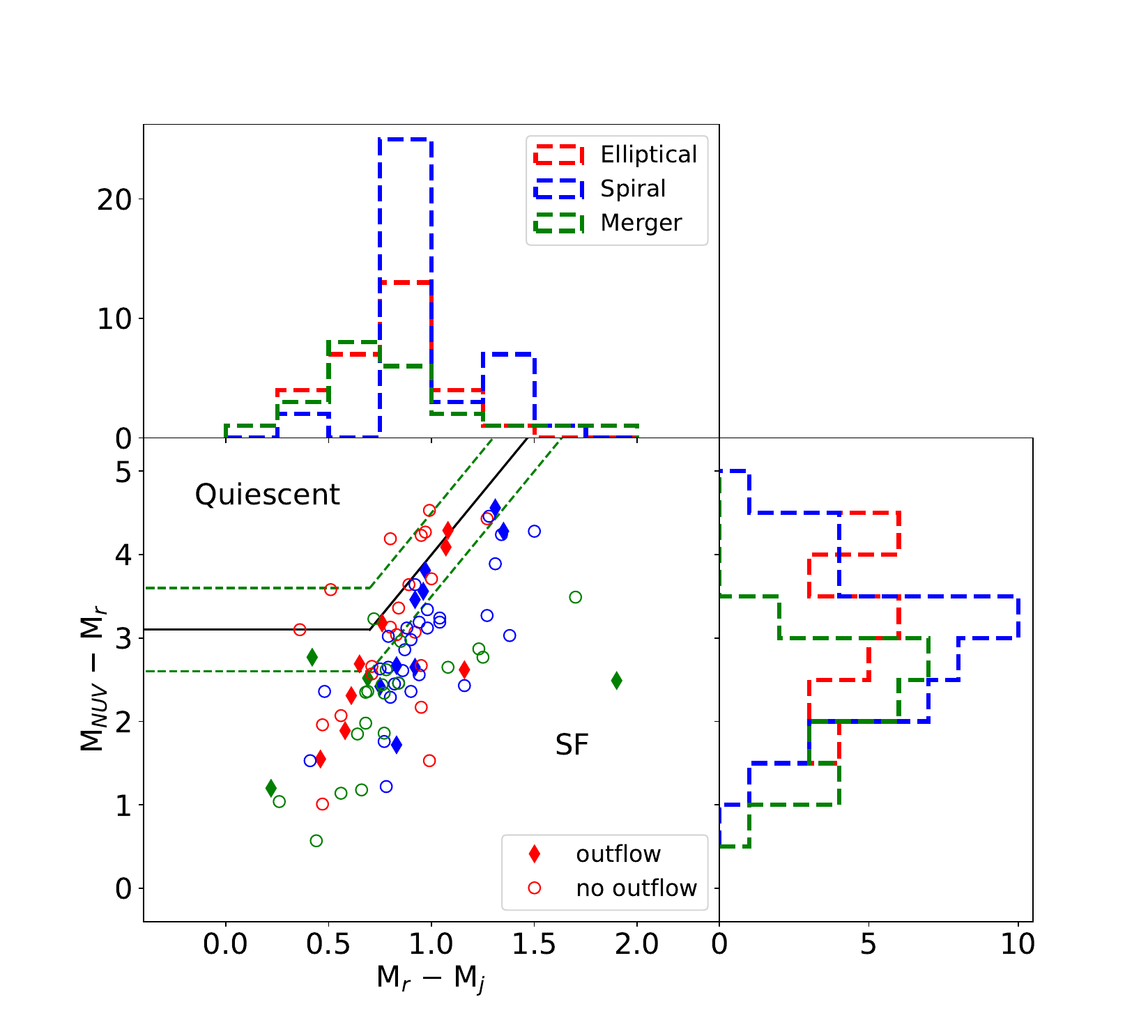}
\caption{Rest-frame (M$_{NUV}$ $-$ M$_r$) vs. (M$_r$ $-$ M$_J$) color-color diagram for galaxies with different morphological types as indicated in the insert. Solid black line mark regions which separates quiescent (upper-left corner) and star-forming galaxies \citep{ilbert13}. Dashed green lines encompass the transitional region know as Green Valley. Filled diamonds represent AGNs with outflow emission. Color distributions are included in the upper and right panels for the objects with different morphological types.}
\label{fig:MNUV_visualclass}
\end{figure}

We revisited the M$_{NUV}$ $-$ M$_{r}$ vs M$_{r}$ $-$ M$_{j}$ color-color diagram as a function of the morphology of their host-galaxy given by the visual classification in Figure \ref{fig:MNUV_visualclass}. We can see that galaxies classified as irregular/mergers show bluer colors with median values of (M$_r$ $-$ M$_j$) = 0.72 and (M$_{NUV}$ $-$ M$_r$) = 2.44, and most galaxies with outflow and merger signatures are found in the SF region (80\%). Spiral galaxies are mostly located in the zone demarcated for SF with 18\% of them in the Green Valley. We see that AGNs with outflow and early type morphology, are evenly located in the GV and SF regions. Noticeably we find 12 (41\%) early type objects with ``blue" colors, populating the demarcated area for star-forming galaxies. 
These objects might belong to the population of galaxies known as \textit{blue ellipticals}, where the star formation is being driven by secular gas accretion processes \citep{lazar2023}.

\begin{figure}
\centering
\includegraphics[width=\columnwidth]{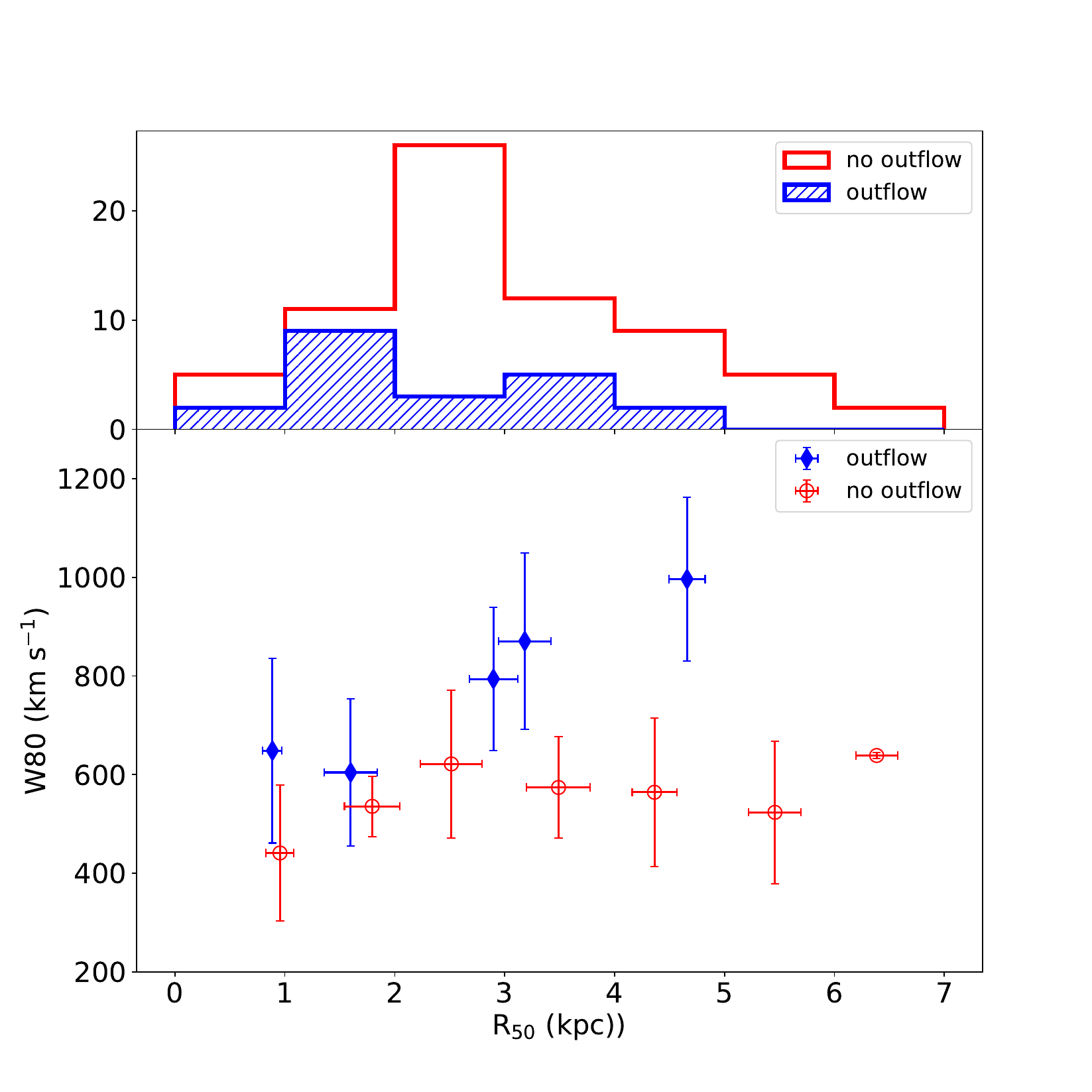}
\caption{Relation between the outflow velocity (W$_{80}$) and the half-light radius.}
\label{fig:W80_R50}
\end{figure}

%When two gaseous galaxies merge, the gas sinks toward the center of the merger product, creating an intense burst of new stars (e.g., Mihos \& Hernquist 1994).
The gas compaction in massive galaxies could also trigger the AGN activity \citep{Ch17,chang2017}. In the compaction scenario \citep{tacchella2016}, galaxies live through one or more blue nugget phases which a minimum in gas depletion time and a maximum in gas fraction are reached.
In Figure \ref{fig:W80_R50} we plot the relation between the outflow velocity (W$_{80}$) and the half-light radius (R$_{50}$). We see a clear positive trend of faster outflow velocity with larger R$_{50}$. Meaning that the more extended (and more massive) galaxies, displays the strongest outflows.
From the distribution of R$_{50}$, on the top panel of Fig. \ref{fig:W80_R50}, we obtain that outflows are preferentially found in AGN with the most compact light distributions. The median half-light radius for galaxies with outflows is R$_{50}$ = 1.8 $\pm$ 1.1 kpc and for galaxies with no outflows R$_{50}$ = 2.7 $\pm$ 1.3 kpc. When segregating according to their morphology, this effect is more noticeable in elliptical galaxies with outflows where their median R$_{50}$ is 1.5 $\pm$ 0.6 kpc, while elliptical galaxies with no outflows have a median R$_{50}$ = 2.0 $\pm$ 0.6 kpc. That is, elliptical galaxies with outflows present a half-light radius of $\sim$80\% the size of AGN of the same morphological class with no outflows.
This is consistent with the scenario proposed by \cite{chang2017} where obscured AGNs are most likely found in star-forming galaxies that have undergone a process of dynamical contraction.
In this scenario, a galaxy's core becomes more compact due to an episode of intense gas inflow, therefore forming a massive bulge with a high gas fraction and star-formation \citep{zolotov2015}. The inflow of gas can also sustain accretion onto the central supermassive black hole, which can trigger the formation of an active galactic nucleus (AGN) with moderately sub-Eddington luminosities. This in turn would favor the formation of the observed AGN-driven outflows.

To further explore how the link between outflows and AGNs is driven, and their impact on their host galaxy, we will need a larger sample of AGNs. This will be achieved thanks to wider AGN surveys, such as those that will be performed by the James Webb Space Telescope (JWST).

\section{Summary}
\label{sec:conclusions}
In this paper, we have studied the galaxy host properties of a sample of AGNs with and without outflow signatures selected on the basis of the MIR color-color diagram proposed by \cite{L04} and the line diagnostic diagram which relates the \oiii/H$\beta$ line ratio and the stellar mass known as mass-excitation diagram \citep{J11,J14}.
We summarize the main results as follows:
 \begin{itemize}
     \item The outflow incidence increases with \oiii~luminosity. Despite this, we do not see a significant trend relating the outflow velocity and the line luminosity.
     \item We also observe a slight trend of higher outflow incidence towards higher X-ray luminosity, with the caveat of having a low statistical significance. The hardness ratio distribution are practically identical for AGNs with and without outflow, together with the previous we can not claim an influence of the inner disc or corona in the triggering of the outflow.
     \item We do not see a significant difference in the host stellar mass distribution nor their star formation rates, between AGNs with outflow signatures and those without them. There is a trend of increasing gas velocities with higher host galaxy mass.
     %\item  There is a hint towards smaller sSFR for AGNs with the fastest outflow velocities. Most likely related to their host galaxy morphological type than due to a quenching process.
     \item By inspecting the M$_{NUV}$ $-$ M$_{r}$ vs. M$_{r}$ $-$ M$_{j}$ color-color diagram, we find that in the Green Valley region there is a higher fraction of AGNs with outflows. The majority of AGNs without outflow ($\sim$75\%) are found in the star forming region.
     \item Morphological analysis from S\'ersic index and non-parametric measurements results in a majority of AGNs with outflows found in galaxies with early type, bulge dominated morphology. From visual inspection we obtain similar fractions ($\sim$ 40\%) of AGN with outflows residing galaxies with disk and elliptical morphological types. We conclude that automated processes to classify the morphological type of a galaxy, fail to correctly identify almost half of the late type galaxies.
     \item It is favored a scenario of dynamical compaction which brings gas into the central part of the galaxy. This would supply the gas necessary to be accreted into the central black-hole, triggering the AGN and the observed outflows.
 \end{itemize}

From our results we can not infer a significant impact on the host galaxy given by the outflow. The feedback claimed by theoretical works is not so evident in our studied sample. AGN feedback if and when present must likely be a local phenomenon, and not galaxy-wide.
On the other hand, large-scale properties of the galaxy such as its mass and morphology, might contribute on the likelihood of observing an AGN-driven outflow.

%% not robust enough to edgeandle our new dual anonymous review requirements and
%% Also note that the akcnowlodgment environment does not support long amounts of text. If you have a lot of people and institutions to acknowledge, do not use this command. Instead, create a new \section{Acknowledgments}.
\section{Acknowledgments}
\begin{acknowledgments}
This work is sponsored by the National Key R\&D Program of China for grant No.\ 2022YFA1605300, 
the National Nature Science Foundation of China (NSFC) grants No. \ 12273051 and 11933003.
This work was partially supported by the Consejo Nacional de Investigaciones Cient\'{\i}ficas y T\'ecnicas (CONICET) and the Secretar\'ia de Ciencia y Tecnolog\'ia de la Universidad Nacional de C\'ordoba (SeCyT). 
Based on data products from observations made with ESO Telescopes at the La Silla Paranal Observatory under ESO programme ID 179.A-2005 and on data products produced by TERAPIX and the Cambridge Astronomy Survey Unit on behalf of the UltraVISTA consortium.
Based on zCOSMOS observations carried out using the Very Large Telescope at the ESO Paranal Observatory under Programme ID: LP175.A-0839 (zCOSMOS).
\end{acknowledgments}

\software{{\tt Astropy} \citep{astropy},
{\tt Matplotlib} \citep{hunter07}, {\sc ifscube} \citep{ifscube}}


%\vspace{5mm}
%\facilities{HST(STIS), Swift(XRT and UVOT), AAVSO, CTIO:1.3m,


%\software{astropy \citep{2013A&A...558A..33A,2018AJ....156..123A},  
%          Cloudy \citep{2013RMxAA..49..137F}, 
%          Source Extractor \citep{1996A&AS..117..393B}
%          }



\appendix
\section{Individual spectral fits}
\label{appendix}

In this Section, we present the spectra with multiple component Gaussian fitting for the sample of 25 galaxies with \oiiil~asymmetric profile chosen by the criteria defined in section \ref{sec:outflow_selection}.

\begin{figure}
    \centering
    \includegraphics[width=\linewidth,height=0.9\textheight]{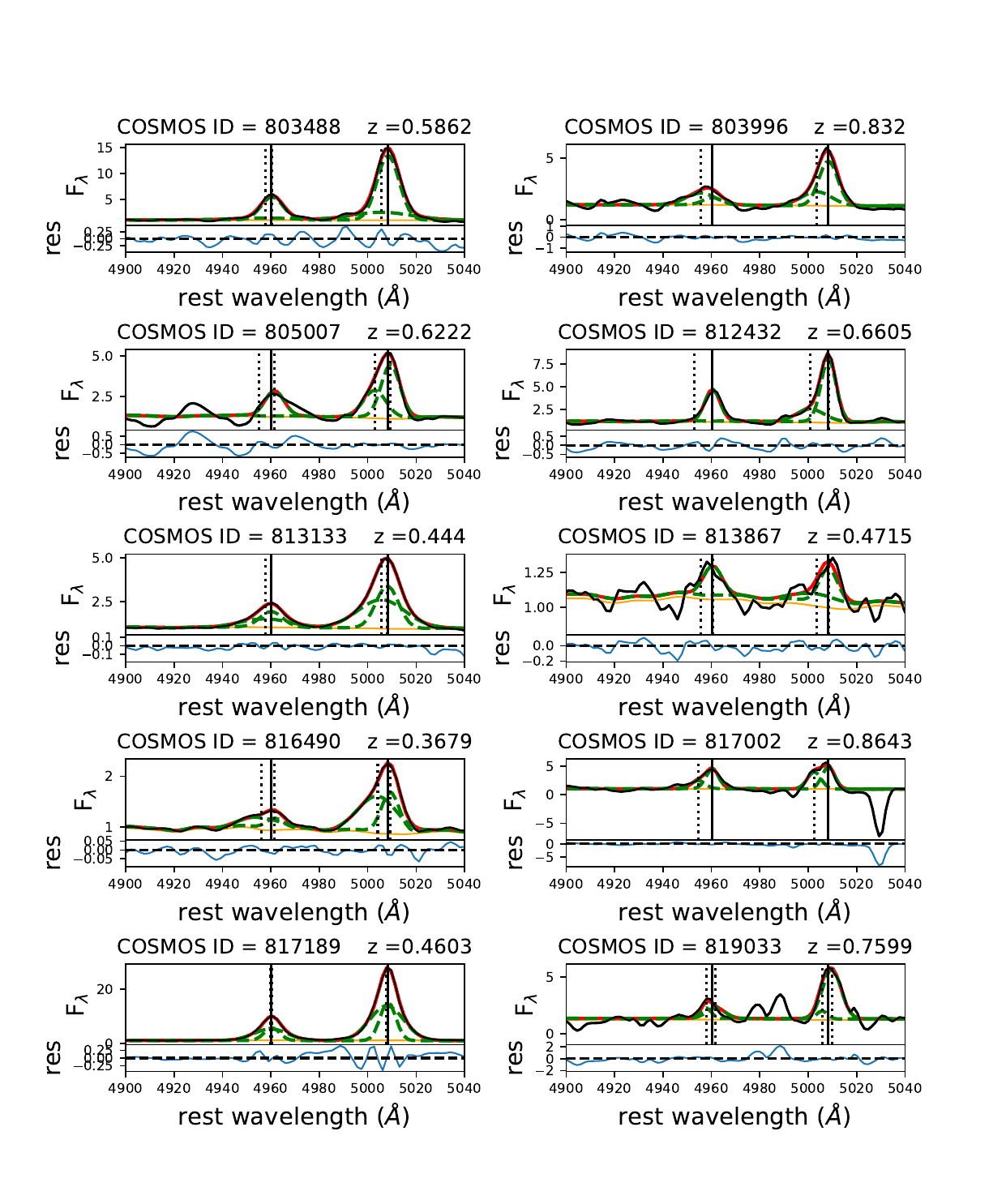}
    \caption{Multiple Gaussian decomposition of the [OIII]$\lambda \lambda$4959,5007 emission lines of the galaxies with asymmetrical [OIII]$\lambda$5007 line profile. In black is shown the observed spectra with its corresponding error, in orange is the stellar continuum determined by \ppxf\, each Gaussian component is shown in slashed green lines, and the full synthetic spectra is plotted in red. The residuals are presented at the bottom panel in cyan. The flux is in units of 10$^{-17}$ erg s$^{-1}$ cm$^{-2}$ \AA{}$^{-1}$.}
    \label{fig:appendix1}
\end{figure}

\begin{figure}
    \centering
    \includegraphics[width=\linewidth,height=0.9\textheight]{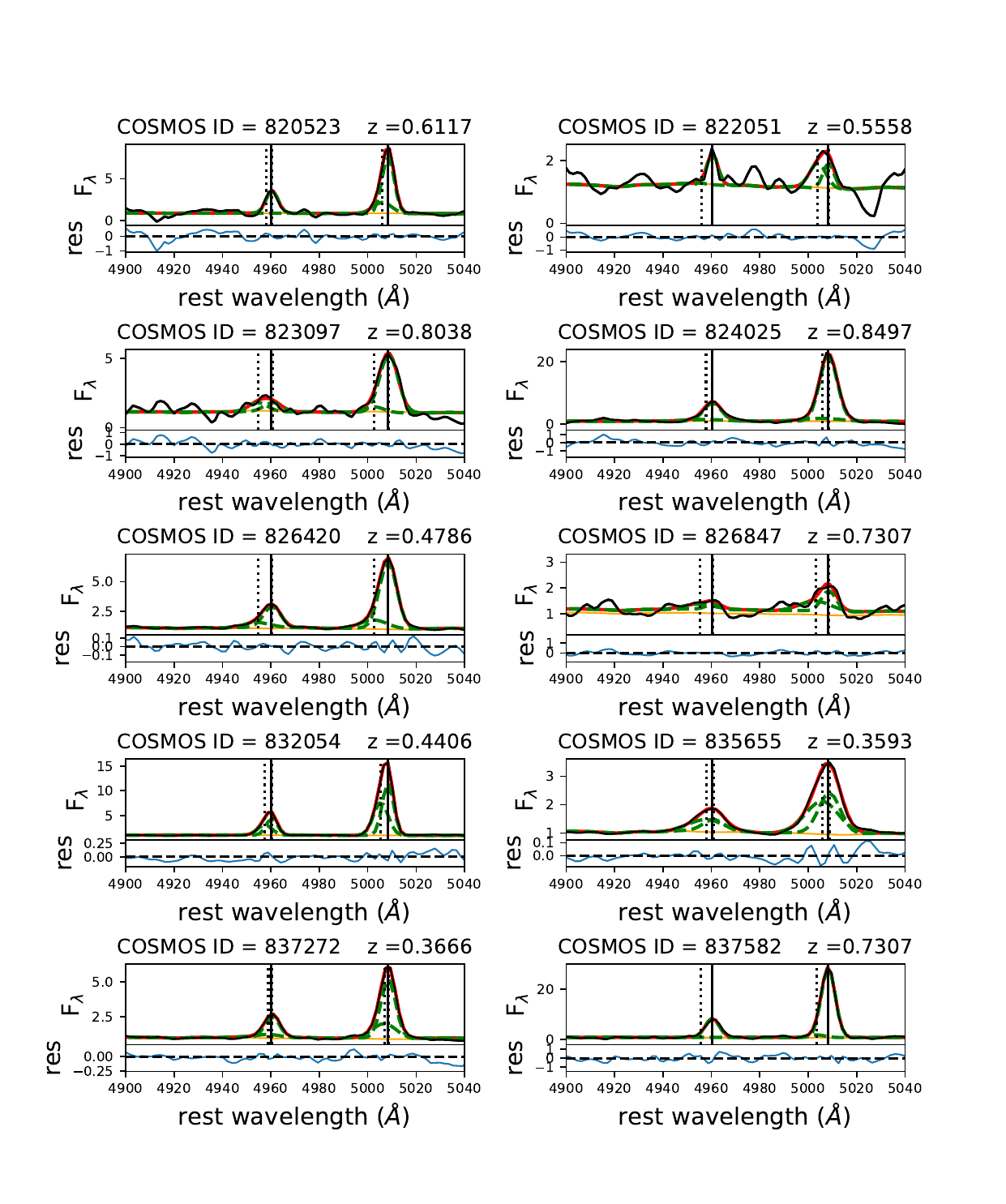}
    \caption{Cont. of Fig. \ref{fig:appendix1}}
    \label{fig:appendix2}
\end{figure}

\begin{figure}
    \centering
    \includegraphics[trim={0 10cm 0 0},clip,width=\linewidth]{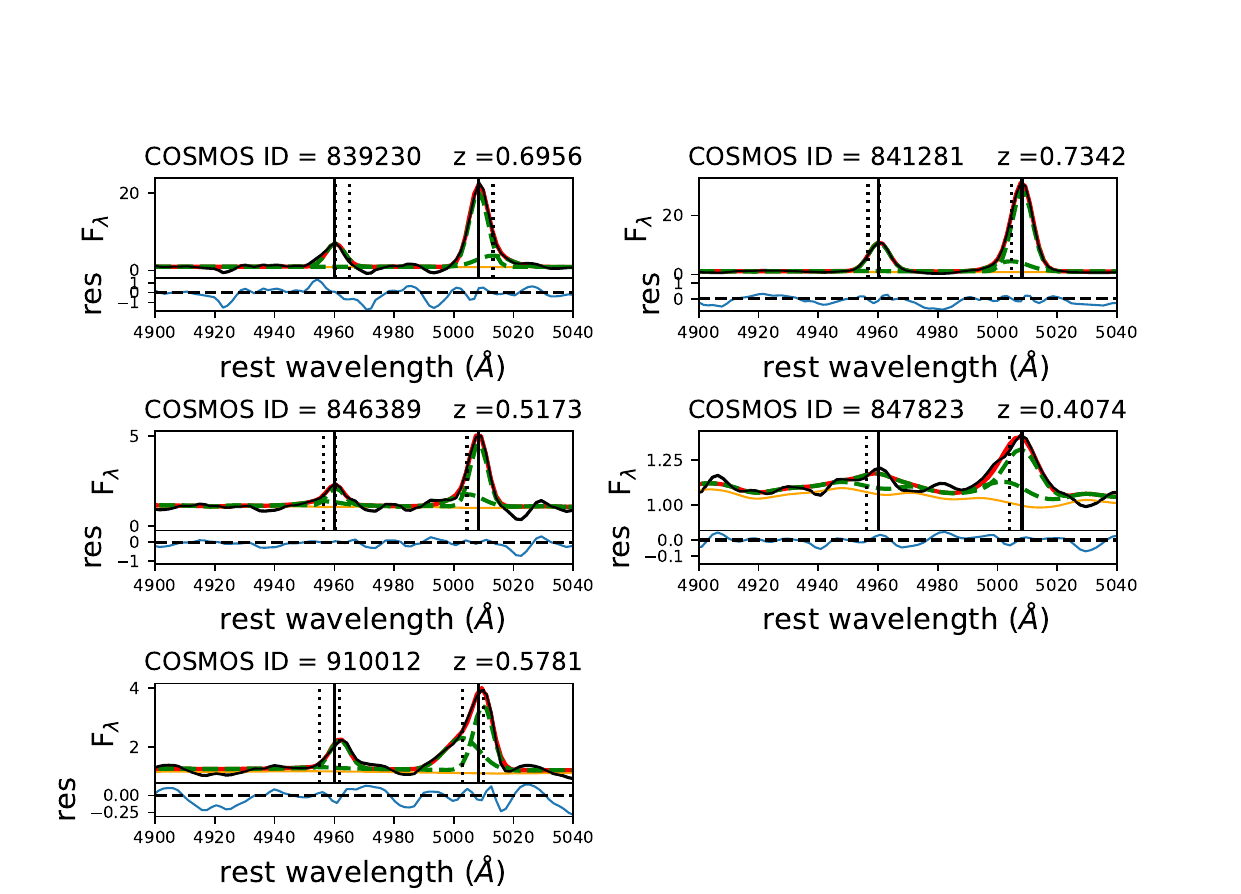}
    \caption{Cont. of Fig. \ref{fig:appendix1}}
    \label{fig:appendix2}
\end{figure}

%\begin{figure}
%    \centering
%    \includegraphics[width=\linewidth,trim={0 11cm 0 0},clip]{gfx/appendix_30_35.pdf}
%    \caption{Cont. of fig. \ref{fig:appendix1}}
%    \label{fig:appendix35}
%\end{figure}

\bibliography{biblio}{}
\bibliographystyle{aasjournal}


\end{CJK*}
\end{document}